\documentclass[onecolumn,superscriptaddress,showpacs,preprintnumbers,amsmath,amssymb,pra]{revtex4}

\usepackage{epsfig}
\usepackage{graphicx}
\usepackage{color}
\usepackage{dcolumn}
\usepackage{bm}

\usepackage{amsmath,amssymb}

\newcommand{\bk}{{{\bf{k}}}}
\newcommand{\bdelta}{{{\bf{\delta}}}}

\newcommand{\bQ}{{{\bf{Q}}}}
\newcommand{\br}{{{\bf{r}}}}

\newcommand{\bS}{{\bf S}}

\newcommand{\dg}{{\dagger}}

\newcommand{\nn}{\nonumber}

\begin{document}

\title{Superfluidity of Dirac Fermions in a Tunable Honeycomb Lattice: \\
Cooper Pairing, Collective Modes, and Critical Currents}

\author{Shunji Tsuchiya}
\affiliation{Department of Physics, Faculty of Science, Tokyo University
of Science, 1-3 Kagurazaka, Shinjuku-ku, Tokyo 162-8601, Japan}
\affiliation{Research and Education Center for Natural Sciences, Keio
University, 4-1-1 Hiyoshi, Kanagawa 223-8521, Japan}
\author{R. Ganesh}
\affiliation{Institute for Theoretical Solid State Physics, IFW Dresden,
Helmholtzstrasse 20, 01069 Dresden, Germany}
\author{Arun Paramekanti}
\affiliation{Department of Physics, University of Toronto, Toronto, Ontario M5S 1A7, Canada}
\affiliation{Canadian Institute for Advanced Research, Toronto, Ontario M5S 1A7, Canada}

\date{\today}

\begin{abstract}
Motivated by recent experiments on atomic Dirac fermions in
a tunable honeycomb optical lattice, we study 
the attractive Hubbard model of superfluidity
in the anisotropic honeycomb lattice. At weak-coupling,
we find that the
maximum mean field pairing transition temperature, as a function of density and 
interaction strength, occurs for the case with isotropic hopping amplitudes. In this isotropic
case, we
go beyond mean field theory and
study collective fluctuations, treating both pairing and density fluctuations
for interaction strengths ranging from weak to strong coupling.
We find evidence for a sharp sound mode, together with a well-defined
Leggett mode over a wide region of the phase diagram.
We also calculate the superfluid order
parameter and collective modes in the presence of nonzero superfluid flow. 
The flow-induced softening of these collective modes leads to
dynamical instabilities involving stripe-like density modulations as well as
a Leggett-mode instability associated with the natural sublattice symmetry
breaking charge-ordered state on the honeycomb lattice. 
The latter provides a non-trivial test for the experimental realization of the one-band Hubbard model.
We delineate regimes of the
phase diagram where the critical current is limited by depairing or by such collective
instabilities, and discuss experimental implications of our results.
\end{abstract}

\maketitle

\section{Introduction}
\label{section1}

Ultracold atoms in optical lattices are of great interest for studying strongly correlated
states of quantum matter, and such emergent phenomena as superconductivity and magnetism \cite{Bloch,Esslinger}.
Stimulated by the discovery of graphene, and ongoing intense research in that field
\cite{CastroNeto}, optical lattices with the honeycomb lattice
structure are beginning to be explored by various groups
\cite{Soltan-Panahi1,Soltan-Panahi2,Tarruell}.
In the context of graphene, various exotic phenomena related to massless Dirac fermions as well as
interesting topological phases and quantum phase transitions have been widely
explored \cite{Haldane,Kane,Kitaev}.
While superconductivity still remains elusive in experiments on graphene, it has been the focus of recent theoretical
work suggesting chiral superconductivity induced by repulsive
interactions \cite{Baskaran,McChesney,Nandkishore,Kiesel}.  Earlier, motivated by a possible cold atom 
realization, where the interatomic interaction may be modelled by an $s$-wave attractive 
contact potential, the attractive fermion Hubbard model on the two-dimensional (2D) honeycomb lattice 
was studied and found to exhibit a quantum phase transition at
half-filling between a semimetal with massless Dirac fermion excitations
and a simple $s$-wave superfluid phase \cite{Zhao,Kopnin1}. This quantum
phase transition was shown to be the end point (with changing density)
of a suitably defined BCS-BEC crossover line away from half filling
\cite{Zhao}. This is reminiscent of the manner in which the usual
BCS-BEC crossover phenomenon in the continuum in 3D at finite density is
linked to the ``zero density quantum critical point''  associated with two-body bound state formation \cite{Nikolic}.
The semimetal to superconductor transition found on the honeycomb lattice is the simplest
version of more general band insulator to superfluid transitions discussed in experiments \cite{Chin}
and theory \cite{Moon,Zhai,Burkov2,Prasad}. The critical theory of this semimetal to
superconductor transition is also a topic of great recent interest \cite{Strack}.
\par
In this paper, motivated by the recent experimental developments reported by the ETH 
group \cite{Tarruell}, we present a careful study of Cooper pairing, superfluid collective modes, 
and superflow instabilities which limit the critical current in the attractive Hubbard model
on the honeycomb lattice. Our main results, which go beyond previous work on this topic,
are as follows. (i) We study the mean field pairing transition temperature $T^0_c$ in a general 
{\it anisotropic} honeycomb lattice as has been realized in a
``brick-wall'' type of geometry
in the ETH experiments \cite{Tarruell}, with hoppings as shown in 
Fig.~\ref{fig.asymcouplings}. For weak coupling,
the maximum $T^0_c$, as a function of fermion density and interaction strength, is found to 
occur in the isotropic limit. We then argue more generally that the vicinity of this isotropic point, 
with all hoppings being equal,
is the most promising limit to experimentally explore the
superfluidity of atomic fermions. (ii) In this limit
of isotropic hopping, we compute 
the collective modes in the superfluid,
treating both density and phase fluctuations on equal footing in contrast to previous
work which only studied superconducting fluctuations \cite{Zhao}. We find evidence for
the expected sharp Anderson-Bogoliubov (AB) sound mode; in addition, over a wide
swath of the phase diagram, we uncover a well-defined Leggett mode, which involves 
intra-unit-cell coupled oscillations of the density and superfluid order parameter phase. 
(iii) Imposing a nonzero superflow, 
we find that the superfluid
phase becomes unstable via either a depairing instability or collective charge modulational
instabilities driven by a softening of the collective mode spectrum. The superfluid is most stable
against depairing in the vicinity of special fillings associated with
van Hove (vH) singularities in the density of states; in this regime, at
mean field level, we find the emergence of a supercurrent induced gapless superfluid state. The collective charge modulational
instabilities that we uncover include tendencies towards forming incommensurate charge orders, 
stripe-like modulations, or sublattice symmetry breaking charge density
wave (CDW) order. Aside from the depairing instability, which occurs
primarily at small fermion density or in the weak attraction limit, we
expect such {\it collective instabilities} to limit the critical current
in this system. 
In fact, the current-induced gapless superfluid state we uncover in
mean field theory is preempted by such collective dynamical or Landau
instabilities for the intermediate coupling regime.
Furthermore, the sublattice CDW order or the instability associated with this CDW
order arises from a subtle pseudospin SU(2) symmetry of the attractive Hubbard model at
half filling; probing for such CDW order 
can thus serve as a strong diagnostic tool to ascertain that the
attractive one band Hubbard model is an appropriate description of the low energy physics 
in a deep optical lattice.
Our predictions can be tested by using Bragg scattering experiments to study 
the collective mode spectrum.
\par
The results described above are obtained using a generalized random phase approximation
(GRPA)
complemented by strong coupling ``pseudospin''-wave theory where appropriate. 
Our work here builds on similar previous studies addressing the collective modes and critical 
current-limiting instabilities on square and cubic optical lattices \cite{Burkov,Ganesh,Yunomae}.
In those cases, it has been shown that the collective mode spectrum of $s$-wave
superfluids exhibits, in addition to
the long wavelength AB sound mode,
a sharp `roton minimum' near the corner of the Brillouin zone (BZ) due
to strong CDW fluctuations, and `roton mode softening' leads to a collective dynamical 
instability of superflow.
\par
This paper is organized as follows.
In Sec.~\ref{sec.anisotropy}, we first examine the honeycomb optical lattice setup from
Ref.~\cite{Tarruell} which can tune the anisotropy of Dirac cones, and discuss the 
mean field $T_c$ as a function of anisotropy $\lambda$, interaction $U$, and filling $n$.
In Sec.~\ref{sec.superflowMFT}, we calculate the superfluid order parameter and quasiparticle 
dispersion in the presence of superflow. 
Section~\ref{sec.collmode} uses the GRPA to compute the spectrum of
collective excitations as a function of flow. In the strong coupling
limit, the GRPA result is compared with a spin wave expansion of the
appropriate pseudospin Hamiltonian. GRPA works well even at strong
coupling, as it shows excellent agreement with spin wave theory.
Finally, Sec.~\ref{sec.instabs} discusses the different kinds of superflow instabilities as a function of density and interaction strength. We find a dynamical instability at the $\Gamma$ point which arises from competing CDW order. In addition, we find an unexpected stripe-like instability which persists even in the strong coupling limit. We conclude by discussing implications for experiments.

\section{Honeycomb lattice with tunable anisotropy}
\label{sec.anisotropy}
The experiments by Tarruell {\it et al}. \cite{Tarruell}, realize a tunable optical 
lattice which could be viewed as a
``brick-wall'' lattice or an anisotropic honeycomb lattice. 
In a deep optical lattice, we can restrict our attention to the lowest band and work with
a single band tight-binding model. On symmetry grounds, two of the
three nearest neighbor hoppings on this lattice are equivalent in the experiments, 
so we denote the three hopping
amplitudes by $(t,t,\lambda t)$ as shown in Fig.~\ref{fig.asymcouplings}
(a).
In the experiments, one pair of sites can be tuned to be closer to each other ($\lambda\geq 1$) or 
further from each other ($0<\lambda < 1$); we therefore carry out a mean field study for
general anisotropy $\lambda > 0$.

\subsection{Attractive Hubbard model}
\begin{figure}
\centering
\includegraphics[width=14cm]{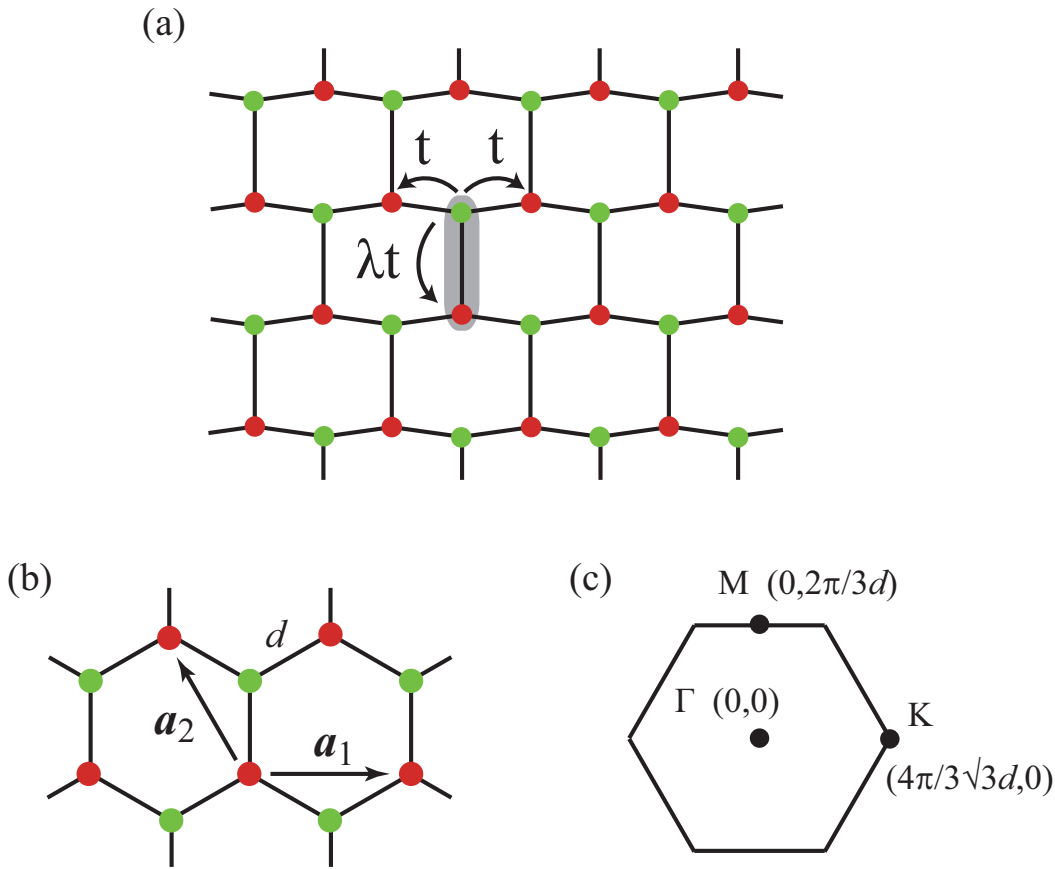}
\caption{(Color online) (a) Honeycomb lattice in a ``brick-wall'' type of
 geometry. Sites of the A(B) sublattice are shown in
 green(red). Asymmetric hopping amplitudes are captured by a parameter
 $\lambda$. $\lambda=1$ is the isotropic limit. The shaded area shows
 the unit cell.  (b) Honeycomb lattice with primitive lattice
 vectors. (c) The Brillouin zone with the high symmetry points
 indicated.
}
\label{fig.asymcouplings}
\end{figure}

Since future experiments
are likely to be able to study atomic fermions with attractive interactions in this lattice, we consider
the attractive Hubbard model on this anisotropic honeycomb lattice as a reasonable starting point
to study the superfluid phase.
The Hamiltonian is thus given by
\begin{eqnarray}
H=-\sum_{\langle i,j\rangle,\sigma}t_{ij} \left(c_{i\sigma}^\dagger c_{j\sigma}+c_{j\sigma}^\dagger
					    c_{i\sigma}\right)
-\mu\sum_{i,\sigma}n_{i\sigma}-U\sum_i n_{i\uparrow}n_{i\downarrow}~,
\label{Hubbard}
\end{eqnarray}
where $c_{j\sigma}^\dagger$ is the creation operator of a fermion
with spin $\sigma$($=\uparrow, \downarrow$) at site $j$, $\langle i,j\rangle$ correspond to nearest neighbor sites coupled by a
hopping amplitude $t_{ij}$, 
the chemical 
potential $\mu$ tunes the fermion density, and $U$ is the local Hubbard attraction.
The hopping $t_{ij} = (t,t,\lambda t)$ is chosen as shown in
Fig.~\ref{fig.asymcouplings} (a) for the three neighboring sites.

\subsection{Single-particle dispersion}
The single-particle Hamiltonian in momentum space takes the form
\begin{equation}
H=\sum_{\bm k,\sigma}\left(\gamma_{\bm k} c_{\bm k, a,\sigma}^\dagger c_{\bm
 k,b,\sigma}+\gamma_{\bm k}^\ast c_{\bm k,b,\sigma}^\dagger c_{\bm
 k,a,\sigma}\right)
 \end{equation}
 where the subscripts $a,b$ refer to the two sublattices, and
\begin{equation}
\gamma_{\bm k}=-t(\lambda+e^{i\bm k\cdot \bm a_2}+e^{i\bm k\cdot (\bm a_1 + \bm a_2)}),
\label{gamma}
\end{equation}
where $\bm a_1=(\sqrt{3}d,0)$ and $\bm a_2=(-\sqrt{3}d/2,3d/2)$ are the 
two basis vectors shown in Fig.~\ref{fig.asymcouplings} (b). This results in two dispersing
bands with energies $\pm |\gamma_{\bm k}|$.
\par
\begin{figure}
\centering
\includegraphics[width=\linewidth]{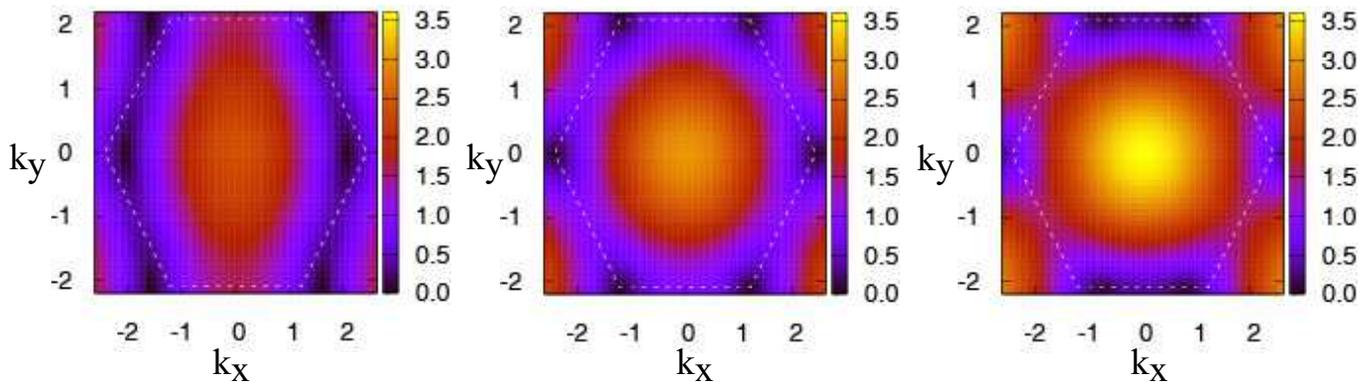}
\caption{(Color online) Top view of dispersion of the non-interacting fermions with $\lambda=0.4$ (left), $\lambda=1$ (center), and $\lambda=1.6$ (right). The white dashed line is the boundary of the first Brillouin zone. In the symmetric $\lambda=1$ limit, the Dirac cones lie at the $K$ points.}
\label{fig.disps}
\end{figure}

For $U=0$, we plot the noninteracting band dispersion $|\gamma_{\bm k}|$
for three different values of $\lambda$ as shown in Fig.~ \ref{fig.disps}.
In the isotropic limit, $\lambda=1$, there are two zero energy Dirac cones which 
occur at the $K$ points (BZ corners) as is familiar from
graphene.
Upon decreasing $\lambda$ from unity, the two Dirac points move towards
the $\Gamma$ point.
When $\lambda \ll 1$, the system becomes quasi one dimensional and the Dirac points
acquire highly anisotropic velocities, with a smaller velocity along the $\hat{y}$ direction.
For $\lambda > 1$, increasing $\lambda$ moves the Dirac cones along the BZ 
edge towards the $M$ point (edge center), with a concomitant tunable velocity anisotropy. 
Eventually, at $\lambda=2$, the two Dirac points meet at the $M$ point and 
`annihilate' each other, resulting in a gapped spectrum
\cite{Zhu,Wunsch,Montambaux}. Such moving and merging/gapping of Dirac
points have been observed in the ETH experiments \cite{Tarruell}.
\par
Figure~\ref{fig.DeltaDOS} shows the density of states (DOS) at the Fermi level
as a function of $\lambda$ and fermion filling
$n$ (here $n=1$ corresponds to `half filling' with one fermion on average per lattice site). 
In the isotropic $\lambda=1$ limit, the DOS exhibits vH
singularities at $n=3/4,5/4$ arising from the hexagonal shape of the
Fermi surface at these densities. Upon decreasing $\lambda$ from unity,
we find that these vH  peaks split into two branches. As we approach the
$\lambda=0$ limit, one of the branches merges with the `band edge' at
$n=0$ or $n=2$. The other branch approaches $n=1$, however it gets
weaker and eventually vanishes at $\lambda=0$. Upon increasing $\lambda$
from unity, we again find that the vH singularities at $n=3/4,5/4$ split
into two branches. For $\lambda \gg 1$, the DOS peaks at densities
$n=0,2$, at energies $\pm \lambda t$ corresponding to the bonding and
antibonding state energies of isolated strong bonds.

\subsection{Cooper pairing: Mean field theory}

\begin{figure}
\centering
\includegraphics[width=\linewidth]{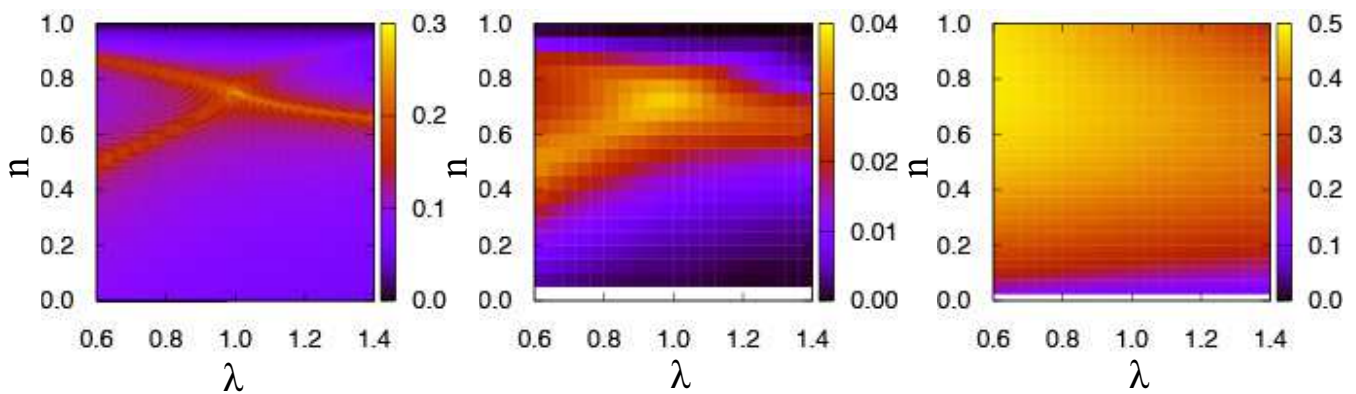}
\caption{(Color online) Left: $g(\epsilon_F)$, density of states at the Fermi level as a function of filling and $\lambda$. We only show $n<1$ as the system is particle-hole symmetric.
Center: Mean field $T_c^0 / t$ with $U/t=1$ as a function of filling (in
 units of fermions per site) and $\lambda$. Right: Mean field $T_c^0 /
 t$ with $U/t=3$ as a function of filling (in units of fermions per
 site) and $\lambda$.} 
\label{fig.DeltaDOS}
\end{figure}

We next consider the effect of the attractive Hubbard interaction. At
mean field level, the ground state of this model is generically a superfluid state formed by condensing Cooper pairs. 
To study this superfluid state, the fermions have to be cooled (at least) below the mean field
transition temperature, $T^0_c$, corresponding to the onset of Cooper pairing. 
We therefore begin by examining $T^0_c$ as a function of anisotropy $\lambda$, and filling. 
\par
At mean field level, the gap and number equations are given by \cite{Zhao}
\begin{equation}
\frac{1}{U}=\frac{1}{N}\sum_{\bm k} \sum_{\tau=\pm} \frac{\tanh\frac{\beta
 E^\tau(\bm k)}{2}}{2E^\tau(\bm k)},
\label{gapeq0}
\end{equation}
and
\begin{eqnarray}
n&=& 1-\frac{1}{N}\sum_{\bm k}\sum_{\tau=\pm}\frac{\xi_{\bm k}^\tau}{E^\tau(\bm
		   k)}\tanh\frac{\beta E^\tau(\bm k)}{2},
\label{number0}
\end{eqnarray}
respectively,
where $\xi^\pm_{\bm k} = \pm |\gamma_{\bm k}| - \mu$ and
$E^\pm(\bm k) = \sqrt{(\xi^\pm_{\bm k})^2 + \Delta^2_0}$ ($\tau$
denotes the energy bands). $N=2M$ is the number of lattice sites and
$M$ is the number of unit cells. The derivation of Eqs.~(\ref{gapeq0})
and (\ref{number0}) is summarized in Appendix A.
To determine $T^0_c$ (and the appropriate $\mu$), we set $\Delta_0 \to 0$ and $\beta \to 1/T^0_c$
in these equations and solve them self-consistently.
\par
Within weak-coupling BCS theory, the pairing gap and $T^0_c$ are determined by 
$g(\epsilon_F)$, which is the DOS at the Fermi level in the non-interacting problem. 
We therefore compare the behavior of the mean field $T^0_c$ with the trends in
$g(\epsilon_F)$. At weak coupling ($U/t=1$),
we find that the maximum $T^0_c$ tracks $g(\epsilon_F)$ for all values of lambda and filling
as seen from Fig.~\ref{fig.DeltaDOS}.
As seen in the figure, the strong vH singularity in the symmetric $\lambda=1$ limit, leads
to particularly robust $T^0_c$ in this isotropic case for $n=3/4,5/4$.
(We will revisit this robustness of the superfluidity in Sec.\ref{subsec.FlowMFT} in the context of imposed superflow.) For intermediate and large $U/t$, the pairing is no longer directly restricted
to electrons near the Fermi surface, and the $T^0_c$ values are no longer governed simply by
$g(\epsilon_F)$ as shown in Fig.~\ref{fig.DeltaDOS}(right). We find that
for \textit{all} interaction strengths, the mean field $T^0_c$ is highest when 
$\lambda \leq 1$ as seen from Fig.~\ref{fig.DeltaDOS}. 
However, as the system becomes increasingly quasi-1D with decreasing 
$\lambda$, we expect enhanced quantum and thermal fluctuations beyond
mean field theory to suppress superfluidity, especially in the strong
coupling regime where the superfluid to normal transition temperature is essentially
controlled by fluctuation effects, and not by pair breaking effects which set
$T^0_c$. On these grounds, we expect the highest superfluid
transition temperature to occur in the vicinity of the isotropic limit $\lambda=1$. We suggest 
$\lambda=1$ and $n=3/4$ (or $5/4$) as the optimal parameter regime 
for studying superfluid order in the experiments.
In the rest of this paper, we therefore set $\lambda=1$ and study the collective
excitations and superflow instabilities of the resulting superfluid state.

\section{Nonzero superflow: Mean field theory on the isotropic honeycomb lattice}
\label{sec.superflowMFT}
We next turn to the mean field theory of the isotropic honeycomb lattice Hubbard model
in the presence of a nonzero superfluid current obtained by considering Cooper pairs with
nonzero center of mass momentum ${\bm Q}$. This will allow us to comment on the
stability of the superfluid towards depairing (pair breaking) effects. This will also set the
stage for the analysis of collective modes and collective mode driven instabilities discussed
in later sections.

\subsection{Mean field theory in the presence of superfluid flow}
\label{BCSMFT}
The order parameter in the presence of superfluid flow
with flow momentum $\bm Q$ is given by
\begin{eqnarray}
\Delta_{\bm Q,\nu}e^{i\bm Q\cdot \bm r_l}&=&U\langle
 c_{l,\nu,\downarrow}c_{l,\nu,\uparrow}\rangle\nonumber\\
&=&\frac{U}{M}\sum_{\bm k}\langle c_{-\bm k+\bm
 Q/2,\nu,\downarrow}c_{\bm k+\bm Q/2,\nu,\uparrow}\rangle e^{i\bm Q\cdot \bm r_l},
\label{gap}
\end{eqnarray}
where $l$ denotes the unit cell and $\nu(=a, b)$ labels the sublattice.
Within the mean field approximation, the Hamiltonian takes the form
\begin{equation}
H=\sum_{\bm k}
\Psi^\dagger_{\bm Q}(\bm k) \hat h_{\bm Q}\Psi_{\bm Q}(\bm
k)+\frac{M}{U}\sum_\nu|\Delta_{\bm Q,\nu}|^2,
\label{Hflow}
\end{equation}
where 
\begin{eqnarray}
\Psi_{\bm Q}(\bm k)=\left(
\begin{array}{c}
c_{\bm k_+,a,\uparrow}\\ c_{\bm k_-,a,\downarrow}^\dagger\\
 c_{\bm k_+,b,\uparrow}\\ c_{\bm k_-,b,\downarrow}^\dagger
\end{array}
\right),~
\hat h_{\bm Q}\equiv 
\left(
\begin{array}{cccc}
-\mu & -\Delta_{\bm Q,a} & \gamma_{\bm k_+} & 0 \\
-\Delta_{\bm Q,a}^\ast & \mu & 0 & -\gamma_{\bm k_-}^\ast\\
\gamma_{\bm k_+}^\ast & 0 & -\mu & -\Delta_{\bm Q,b}\\
0 & -\gamma_{\bm k_-} & -\Delta_{\bm Q,b}^\ast & \mu
\end{array}
\right).
\label{MFTHmlt}
\end{eqnarray}
Here, $\bm k_\pm$ denotes $\pm\bm k+\bm Q/2$.
The matrix $\hat h_{\bm Q}$ can be diagonalized by the Bogoliubov transformation
\begin{equation}
\Psi_{\bm Q}(\bm k)=\hat T\tilde\Psi_{\bm Q}(\bm k)=
\hat T
\left(
\begin{array}{c}
\alpha_{\bm k,+,\uparrow}\\
\alpha_{-\bm k,+,\downarrow}^\dagger\\
\alpha_{\bm k,-,\uparrow}\\
\alpha_{-\bm k,-,\downarrow}^\dagger
\end{array}
\right),
\label{BogTrans}
\end{equation}
where $\alpha_{\bm k,\tau,\sigma}^\dagger$ is the creation operator of a
Bogoliubov quasiparticle with momentum $\bm k+\bm Q/2$, energy band
$\tau$$(=\pm)$, and spin $\sigma$$(=\uparrow,\downarrow)$. They 
satisfy the anticommutation relation $\{\alpha_{\bm
k,\tau,\sigma},\alpha^\dagger_{\bm k',\tau',\sigma'}\}=\delta_{\bm 
k,\bm k'}\delta_{\tau,\tau'}\delta_{\sigma,\sigma'}$.
The unitary matrix $\hat T$ is determined so as to diagonalize the mean field Hamiltonian
\begin{eqnarray}
H'&=&\sum_{\bm k}\Psi^\dagger_{\bm Q}(\bm k) \hat h_{\bm Q} \Psi_{\bm
 Q}(\bm k)=\sum_{\bm k}\tilde\Psi^\dagger_{\bm Q}(\bm k)\hat{\tilde h}_{\bm Q}\tilde\Psi_{\bm Q}(\bm k),\\
\hat{\tilde h}_{\bm Q}
&=&\hat T^\dagger \hat h_{\bm Q} \hat T=
\left(
\begin{array}{cccc}
E^+_{\bm Q}(\bm k) & 0 & 0 & 0\\
0 & -E^+_{\bm Q}(-\bm k) & 0 & 0\\
0 & 0 & E^-_{\bm Q}(\bm k) & 0\\
0 & 0 & 0 & -E^-_{\bm Q}(-\bm k) \\
\end{array}
\right),
\end{eqnarray}
where $E^+_{\bm Q}(\bm k)$ and $E^-_{\bm Q}(\bm k)$ are the upper and
lower quasiparticle energy bands, respectively.
Thus, we obtain the diagonalized Hamiltonian
\begin{equation}
H=\sum_{\bm k}\sum_{\tau=\pm}\left[E^\tau_{\bm Q}(\bm k)(\alpha_{\bm k,\tau,\uparrow}^\dagger\alpha_{\bm k,\tau,\uparrow}+\alpha_{\bm k,\tau,\downarrow}^\dagger\alpha_{\bm k,\tau,\downarrow})-E^\tau_{\bm Q}(\bm
	       k)\right]+\frac{M}{U}\sum_\nu|\Delta_{\bm
	       Q,\nu}|^2,
\end{equation}
and the ground state energy 
\begin{equation}
E_0=-\sum_{\bm k}\sum_{\tau=\pm}E^\tau_{\bm Q}(\bm
					  k)+\frac{M}{U}\sum_{\nu}|\Delta_{\bm Q,\nu}|^2.
\end{equation}
To evaluate $\Delta_{\bm Q,a/b}$ and $\mu$, we solve the corresponding gap and number equations that
are explicitly given in Appendix \ref{app.MFTeqs}. The chemical
potential is tuned to obtain the required density and the order
parameters are chosen to minimize the ground state energy.
This scheme is known to interpolate between the weak-coupling BCS regime and strong-coupling BEC regime at low
temperatures \cite{Eagles,Leggett,NSR}.
The strong-coupling effect is included in the self-consistently
determined chemical potential, which reduces to the Fermi energy in the
weak-coupling BCS regime while it deviates from the Fermi energy in the strong coupling
BEC regime.
\par
For small order parameter and very small $\bm Q$, we find a current in mean field theory
which is linear in $\bm Q$ at all densities, including 
at and near half filling where a linearized Dirac theory
\cite{Kopnin1,Kopnin2} would be applicable. However, our
reported results mostly explore the regimes of large superflow momenta
and a wide range of fillings, where a linearized
Dirac dispersion is not applicable and the mean field renormalization of the
gap must also be taken into account. Furthermore, as we show in the next section, once we
consider physics beyond mean field theory, 
any nonzero current is unstable at half-filling for the attractive
Hubbard model due to a dynamical instability associated 
with charge order.

\subsection{Depairing instability and gapless superfluidity}
\label{subsec.FlowMFT}

\begin{figure}
\centerline{\includegraphics[width=10cm]{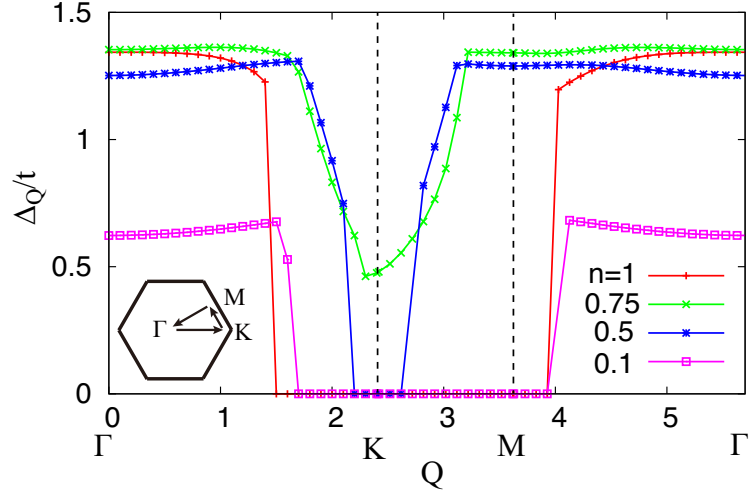}}
\caption{(Color online) Superfluid order parameter $\Delta_{\bm
 Q}(=\vert\Delta_{\bm Q,a}\vert=\vert\Delta_{\bm Q,b}\vert)$ as a function of
 the superflow $\bm Q$ for several different fermion fillings at $T=0$
 for $U/t=4$.
 $\bm Q$ is swept along the contour displayed in the inset.}
\label{delta_Um40}
\end{figure}
Figure~\ref{delta_Um40} shows the self-consistently calculated order
parameter as a function of superflow momentum $\bm Q$ for different
fillings. Hereafter in this section and in Sec.~\ref{sec.collmode}, we
restrict ourselves at $T=0$ for simplicity.
For any flow direction, we find that the amplitudes of the two order
parameters are the same, i.e., $\vert \Delta_{\bm Q, A}\vert = \vert
\Delta_{\bm Q, B}\vert\equiv\Delta_{\bm Q}$.
We plot this amplitude $\Delta_{\bm Q}$ as a function of $\bm Q$ for different fillings in Fig.\ref{delta_Um40}.
Since the Hamiltonian is particle-hole symmetric, we
restrict ourselves to densities below half-filling($n\leq 1$).
Sweeping $\bm Q$ in the BZ,
we find that the order parameter generally vanishes
discontinuously at some critical flow momentum (except in the vicinity of $n=3/4$).
This depairing instability occurs when the kinetic energy cost of superflow outweighs the 
energy gain from pairing. The system then reverts to an unpaired normal state
via a first order phase transition.
\par
However, when the filling is close to $n=3/4$ fermions per site, $\Delta_{\bm Q}$ 
does not vanish for any $\bm Q$. This enhanced stability is due to a vH singularity in the non-interacting problem at $n=3/4$ and $n=5/4$ (see Fig.\ref{fig.DeltaDOS}). At these densities, there is a very large density of states at the Fermi level which enhances superfluid order and makes it robust against superflow. 
\par
\begin{figure}
\centerline{\includegraphics[width=15cm]{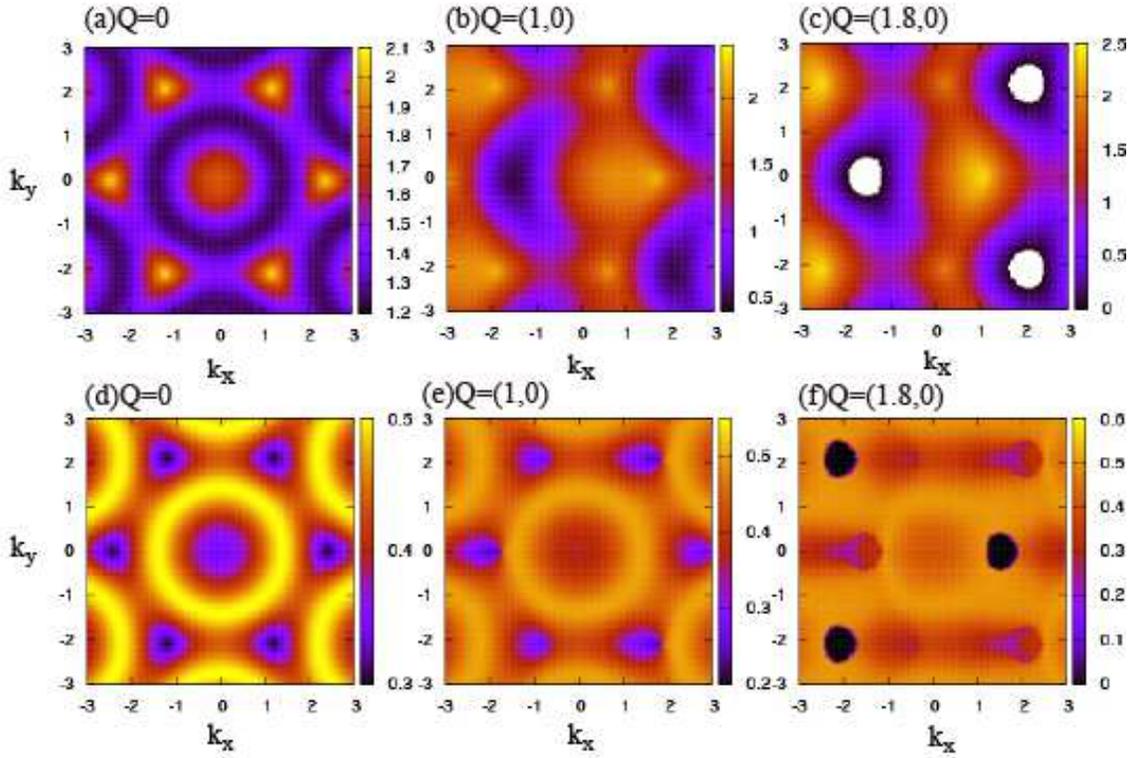}}
\caption{(Color online) ((a)$\sim$(c)) Single-particle spectrum $E^-_{\bm Q}(\bm k)$
 (in units of $t$) and ((d)$\sim$(f)) order parameter weight $w^-_{\bm
 Q,A}(\bm k)$ (in units of $t$) in the lower 
 excitation energy band for different flow momenta $\bm Q$. We set
 $U/t=4$, $T=0$, and $n=0.5$. In panel (c), $E^-_{\bm Q}(\bm k)$ takes negative values in 
 the white regions indicating the gapless superfluid state.
}
\label{uvplot}
\end{figure}
In Fig.~\ref{delta_Um40}, each curve for $\Delta_{\bm Q}$ exhibits an interesting
{\it cusp} structure, at which the derivative $\partial\Delta_{\bm
Q}/\partial \bm Q$ changes discontinuously. The order parameter starts
decreasing rapidly beyond the cusp until the transition to the normal
state sets in. This cusp structure signals the onset of {\it gapless superfluidity}. 
The evolution of the single-particle spectrum in the
lower band $E^-_{\bm Q}(\bm k)$ is shown in Figs.~\ref{uvplot} (a)$\sim$(c).
With increasing flow, the energy gap to creating quasiparticles decreases 
for wavevectors $\bm k$ in the direction opposite to $\bm Q$
due to a `Doppler shift' in the dispersion.
For large $\bm Q$, the gap closes making the single-particle spectrum gapless -
 this precisely coincides with the cusp feature.
However, the closing of the gap does not immediately lead to
depairing, as the self-consistent order parameter remains non-zero.
\par
To clarify the origin of this cusp structure, we plot the
single-particle spectrum of the lower energy band $E^-_{\bm Q}(\bm k)$,
as well as the order parameter weight for the same band defined as
\begin{equation}
w^-_{\bm Q,a}(\bm k)\equiv T_{23}^*T_{13}f(E_{\bm Q}^-(\bm
 k))+T_{24}^*T_{14}(1-f(E_{\bm Q}^-(-\bm k))).
\label{weight}
\end{equation}
Here, $f(E)\equiv 1/(e^{\beta E}+1)$ is the Fermi distribution function
and $T_{ij}$ denotes the $ij^{th}$ element of the Bogoliubov
transformation matrix in Eq.~(\ref{BogTrans}).
This quantity $w^-_{\bm Q,a}(\bm k)$ is the contribution from wavevector $\bm k$ 
in the gap equation, Eq.~(\ref{gapA}).
As the quasiparticles are typically gapped, $f(E_{\bm Q}^-(\bm k))=0$ at
$T=0$ and the only contribution to the weight 
comes from the term proportional to $(1-f(E_{\bm Q}^-(-\bm k)))$.
If the quasiparticle energy becomes negative for a certain momentum $\bm p$, i.e.,
$E^-_{\bm Q}(\bm p)<0$, we have $f(E^-_{\bm Q}(\bm p))=1$. 
Now, the order parameter weight at $\bm k = -\bm p$, as defined in Eq.~(\ref{weight}), vanishes as 
$(1-f(E_{\bm Q}^-(-\bm k)))=0$. This is clearly seen in Fig.~\ref{uvplot} (c) and (f).
The single-particle spectrum becomes gapless at some wavevectors $\bm p$, and 
the order parameter weight vanishes at $-\bm p$. 
Thus, the closing of the single-particle gap suppresses the order parameter weight at certain pockets. Consequently, the 
order parameter subsequently decreases sharply, but remains
non-zero. This is the origin of the cusp in $\Delta$ versus $\bm Q$ at the
onset of gapless superfluidity.

\section{Collective mode analysis}
\label{sec.collmode}

The mean field theory in Sec.~\ref{BCSMFT} describes a superfluid
state where Cooper pairs condense in a single momentum state with uniform density. 
To investigate the stability of superfluid order with imposed flow, we
study quantum fluctuations beyond mean field theory. These fluctuations
give rise to dispersing collective modes in the superfluid. For
concreteness, we restrict ourselves to flow in the $\Gamma-K$ direction
in the rest of this paper.
This direction is best suited for the optical lattice setup of
Ref.~\onlinecite{Soltan-Panahi1} which uses three lasers at angles of
120$^\circ$ with respect to each other. Flow can easily be induced in
the $\Gamma-K$ direction by detuning one of the lattice lasers.
For concreteness, we take flow to be parallel to the ${\bm a}_1$ lattice vector (see Fig.\ref{fig.asymcouplings}b). Together with our definition of the unit cell, this forces $\Delta_{\bm Q,a} = \Delta_{\bm Q,b}$ by symmetry. 
\par
In general, the collective mode spectrum can be evaluated from the poles of the
relevant response functions. 
In the superfluid state, we have fluctuations in the superfluid order parameter
which are coupled to fluctuations in density. Therefore, we
calculate the density response function as well as the pair response function. 
We use the GRPA which treats density and pairing fluctuations on an
equal footing. In the context of ultracold Fermi superfluids, GRPA has
successfully explained \cite{Hu} Bragg scattering measurements \cite{Veeravalli} in the absence of an optical lattice.
Various formulations of this method have been used to study the collective modes of
Fermi superfluids in deep square/cubic optical
lattices \cite{Ganesh,Yunomae,Koinov}.
Although GRPA is set up as a weak-coupling perturbation approach, it
successfully captures the strong coupling limit as well.
\par
In this paper, we present GRPA results based on the Green's
function approach developed by C\^ot\'e and Griffin \cite{Cote}. For the case of
square/cubic lattice, this formulation \cite{Yunomae} is in good quantitative agreement
with a perturbation theory method used by two of us earlier \cite{Ganesh}. For the honeycomb lattice case, we have explicitly checked that these two formulations are also in agreement. 
We briefly summarize the C\^ot\'e-Griffin formalism in Appendix B.
At strong coupling, we compare GRPA results with spin wave analysis of the appropriate pseudospin Hamiltonian. The formalism of the spin-wave analysis is summarized in Appendix C.

\subsection{Collective modes in the stationary ground state}

\begin{figure}
\centerline{\includegraphics[width=\linewidth]{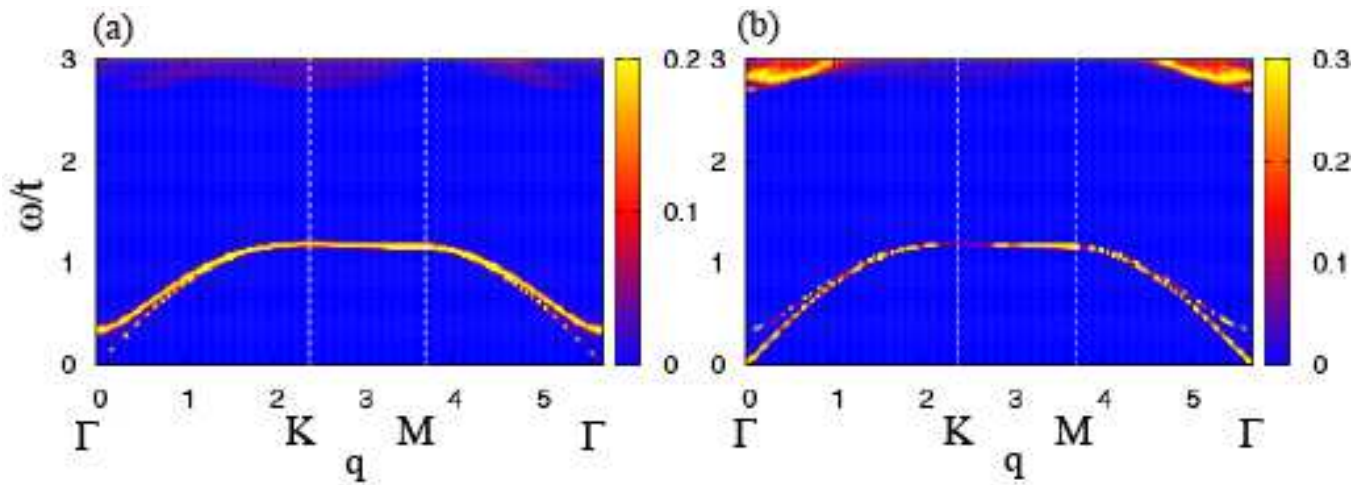}}
\caption{(Color online) Calculated intensity of the (a) dynamic structure
 factor of density reponse function ($S^{d}(\bm q,\omega)$)
 and (b) that of pair response function ($S^{\Delta}(\bm q,\omega)$)
 in the ground state without flow ($\bm Q=0$) for $U/t=4$, $T=0$, and $n=0.9$. 
The momentum $\bm q$ is along the contour displayed in the inset of Fig.~\ref{comparison}.
 The gapped Leggett mode (see text) has a large density component, while the
 gapless AB mode has a large pairing component.
}
\label{chinm}
\end{figure}
\begin{figure}
\centerline{\includegraphics[width=8cm]{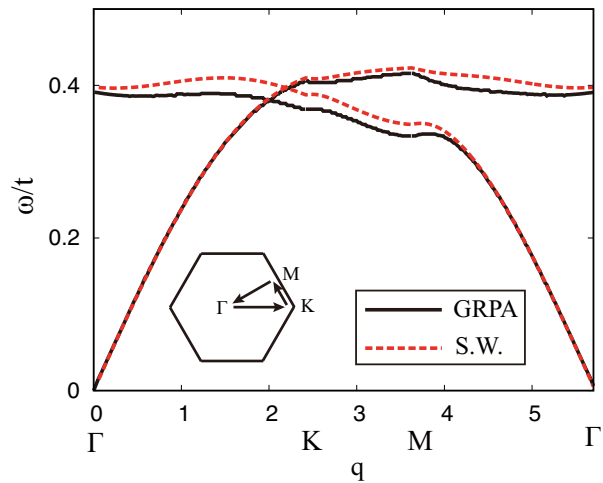}}
\caption{(Color online) Comparison of the collective mode energy in strong coupling regime
 ($U/t=15$) along the contour displayed in the inset. We set $n=0.5$, $T=0$, and
 $\bm Q=(0.2,0)$.
The GRPA result (solid line) is in good agreement with the spin-wave (S.W.) result
 (dashed line) for the strong-coupling pseudospin model. 
}
\label{comparison}
\end{figure}
\begin{figure}
\centerline{\includegraphics[width=\linewidth]{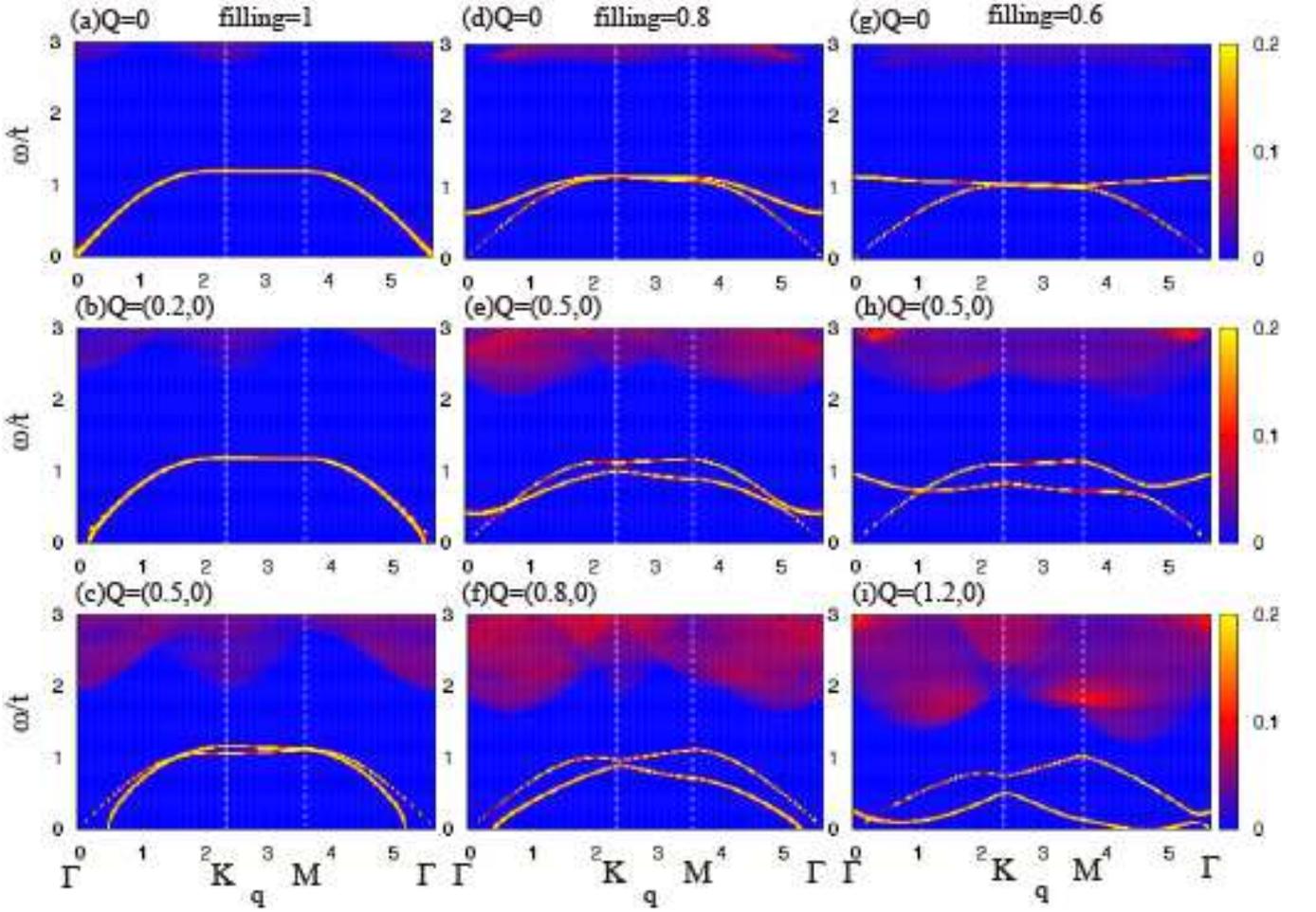}}
\caption{(Color online) Calculated intensity of the dynamic structure
 factor of density response function ($S^{d}(\bm q,\omega)$) for
 different fillings and flow momenta, at $U/t=4$ and $T=0$.
The momentum $\bm q$ is along the contour displayed in the inset of Fig.~\ref{comparison}.
The Leggett mode goes down in energy with increasing flow and `softens', signaling an instability.
}
\label{chin_Um40_1}
\end{figure}
\begin{figure}
\centerline{\includegraphics[width=\linewidth]{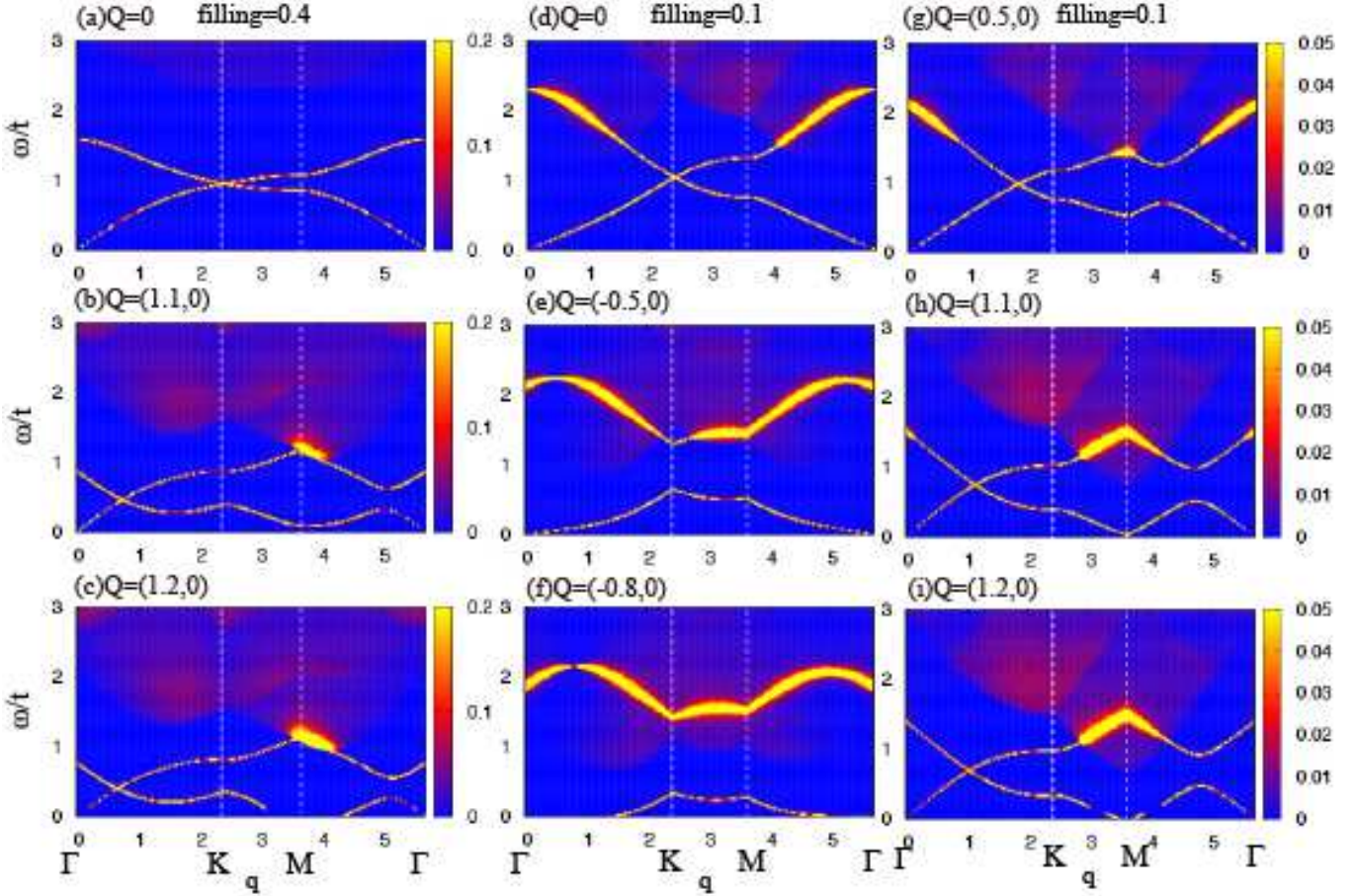}}
\caption{(Color online) Calculated intensity of the dynamic structure
 factor of density response function ($S^{d}(\bm q,\omega)$) for
 different fillings and flow momenta, at $U/t=4$ and $T=0$.
The momentum $\bm q$ is along the contour displayed in the inset of Fig.~\ref{comparison}.
The Leggett mode undergoes a dynamical instability at the $M$ point for
 $n=0.4$. For $n=0.1$, the usual Landau instability induced by long wavelength AB phonons occurs.
}
\label{chin_Um40_2}
\end{figure}
Figures~\ref{chinm}$\sim$\ref{chin_Um40_2} show the dynamic structure factor for density response
and pairing response for different values of filling and
flow. They are defined, respectively, as 
\begin{eqnarray}
S^{d}(\bm q,\omega)&=&-\frac{1}{\pi}{\rm Im}\chi_d^{\nu\nu}(\bm
 q,i\Omega_n\to\omega+i\delta),\\
S^{\Delta}(\bm q,\omega)&=&-\frac{1}{\pi}{\rm Im}\chi_\Delta^{\nu\nu}(\bm q,i\Omega_n\to\omega+i\delta).
\end{eqnarray}
Details are given in the Appendix B.
The collective mode spectrum appears as a sharp peak in intensity of
either dynamic structure factor, while the broad low intensity region
is the two particle continuum.
\par
Since the GRPA equations of motion for the density and pair correlation functions have
the same diagrammatic structure (see Eqs.~(\ref{GRPAladder}),
(\ref{GRPAbubble}), and Fig.~\ref{GRPA_diagram}), they share the same
poles and therefore the same collective mode spectrum. However, the spectral weight of the poles
can be different for density and pairing correlation functions.
The physical origin of collective modes can be understood from the relative weights for density and pairing response.
To illustrate this, we discuss collective modes in the stationary ground state ($\bm
Q=0$) in this section. 
\par
The honeycomb lattice superfluid can have two undamped collective modes due to the presence of two sites in the unit cell, as shown in Fig.~\ref{chinm}: \\
(i) The lower mode incorporates the gapless Nambu-Goldstone mode corresponding to U(1) symmetry breaking. It is
the usual `Anderson-Bogoliubov' (AB) mode, predominantly composed of in-phase pairing fluctuations. 
This is seen from Fig.~\ref{chinm}, where close to the $\Gamma$ point, the AB mode shows a large intensity of the dynamic structure factor for pairing response. \\
(ii) The upper mode is a gapped `optical' branch arising from density and/or phase
fluctuations which are out of phase between two sublattices.
This optical branch can be thought of as a `Leggett mode' \cite{Leggett2}, in analogy with
two-band superconductors such as MgB$_2$ \cite{Blumberg}.
When the interaction strength is not too large, as shown in
Fig.~\ref{chin_Um40_2} (d), the Leggett mode may enter the two particle
continuum and be strongly damped \cite{Zhao}.
\par
Precisely at half-filling, the Hubbard model possesses a special SU(2)
pseudospin symmetry \cite{Zhang}. Consequently, the superfluid ground state is degenerate with a CDW state which breaks sublattice symmetry. As a result, the Leggett mode which is composed of out-of-phase density fluctuations, becomes gapless. We are left with two Nambu-Goldstone modes, as the superfluid state breaks pseudospin SU(2) symmetry and not just U(1) gauge symmetry.
Tuning away from half-filling, this degeneracy is weakly broken. The
Leggett mode acquires a small gap as shown in
Figs.~\ref{chin_Um40_1} (a)-(d)-(g). Close to half-filling, we deduce that the
Leggett mode arises from low-lying CDW fluctuations - as shown in
Fig.~\ref{chinm}, this mode has much larger intensity for density
response than for pairing. Far away from half-filling, CDW fluctuations
are suppressed and both collective mode branches are predominantly
composed of superfluid phase fluctuations.  
\par
In addition to these collective modes, there exists a superfluid amplitude mode above
the two particle continuum; it appears as a large peak at the edge of
the continuum in Fig.~\ref{chinm} (b). 
However, this amplitude mode is strongly damped by decay into
pairs of single-particle excitations. Consequently, it may prove
difficult to detect in experiments. We note that the amplitude mode very weakly couples
with density fluctuations, if at all. As a consequence, the density correlation function does not
show a peak at the edge of the continuum in Fig.~\ref{chinm} (a).

\subsection{Superflow instabilities}
\label{sec.instabs}
As described in Sec.~\ref{BCSMFT}, we impose superfluid flow on the system by
forcing pairing at finite momentum.
We use the GRPA method to extract the collective mode as a function of
flow. As discussed earlier, GRPA also captures the strong coupling limit
as shown in Fig.~\ref{comparison} by comparison with strong-coupling spin wave
expansion. This gives us confidence that GRPA can be used to study
superflow instabilities at all interaction strengths.
Superfluidity can be sustained until some critical flow momentum which is
set by one of three possible mechanisms \cite{Ganesh}:\\
(I)\ Depairing instability: the mean field superfluid order parameter 
vanishes. As discussed in Sec.~\ref{subsec.FlowMFT}, this instability occurs
at the mean field level for weak to intermediate coupling
strengths. \\
(I\hspace{-.1em}I) Landau instability: the collective mode energy hits zero and
subsequently becomes negative \cite{Lifshitz}. The negative-energy modes are populated
after the onset of the instability, leading to dissipation and the loss of superfluidity.
As in the uniform superfluid Fermi gas \cite{Miller}, this instability dominates in the
strong-coupling BEC regime.\\
(I\hspace{-.1em}I\hspace{-.1em}I) Dynamical instability: the
collective mode dispersion hits zero and 
subsequently becomes complex-valued. At the onset of this instability, one or more fluctuation modes start growing exponentially. Due to the rapid growth of fluctuations,  this instability should be easy to observe in experiments. 
In fact, dynamical instabilities in bosonic condensates in a 1D optical
lattice have been successfully observed \cite{Fallani}.
\par
\begin{figure}
\centerline{\includegraphics[width=10cm]{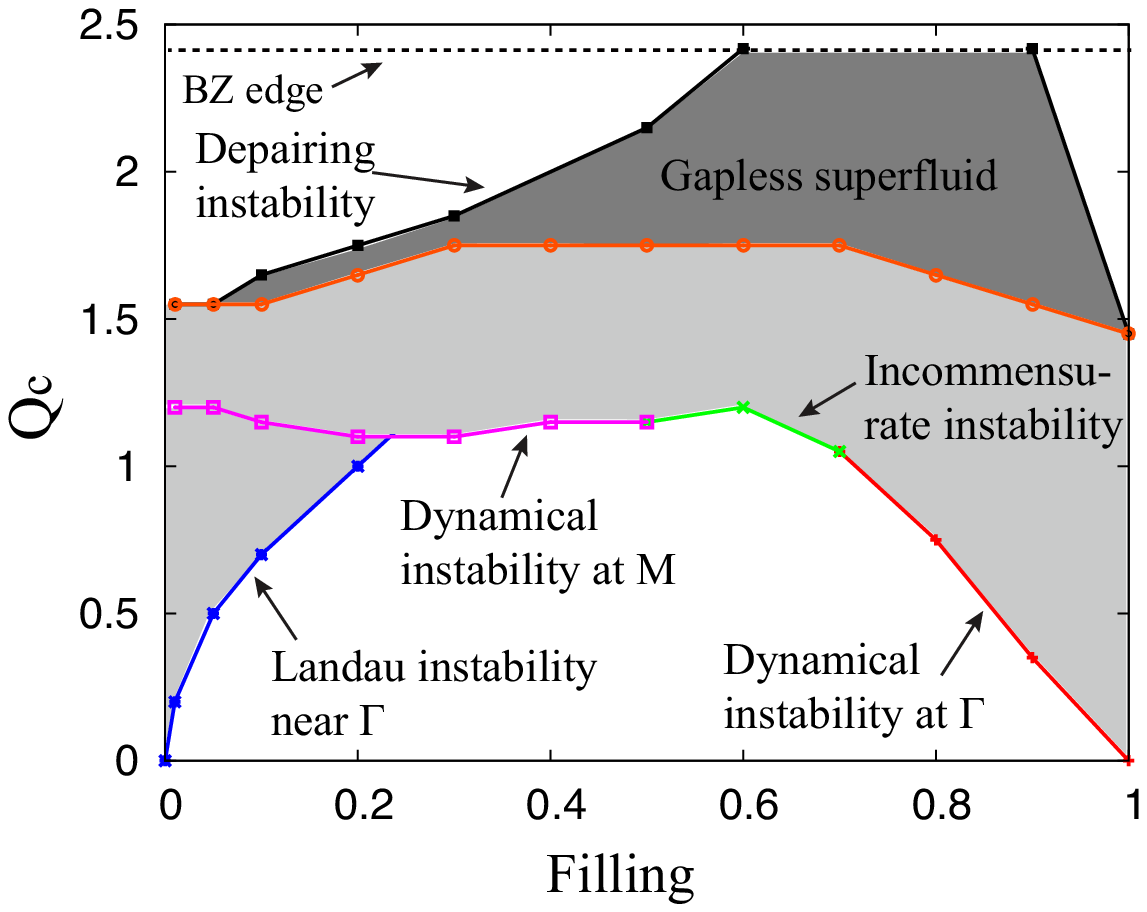}}
\caption{(Color online) Stability phase diagram for $U/t=4$. The magnitude of the
 critical momentum is plotted as a function of flow imposed along the
 $x$-axis ($\Gamma-K$ direction). In the light gray region, superfluidity is
 destroyed by collective mode instabilities despite the finite
 single-particle energy gap. The dark gray region indicates the gapless
 superfluid region (see text).
 Note that the depairing instability does not occur for
 $0.6\leq n\leq 0.9$ due to the robust superfluidity in the vicinity of
 the vH singularity at $n=3/4$.
}
\label{PD_U40}
\end{figure}
\begin{figure}
\centerline{\includegraphics[width=15cm]{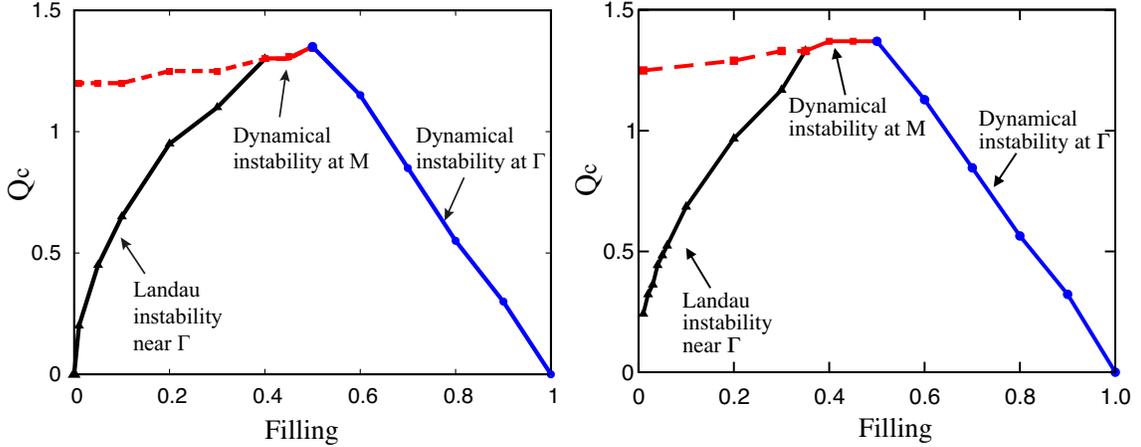}}
\caption{(Color online) Comparison of the stability phase diagram
 obtained by the GRPA (left) and spin-wave theory (right) in the
 strong-coupling regime $U/t=8$. The magnitude of the
 critical momentum is plotted against flow imposed along the
 $x$-axis ($\Gamma-K$ direction).
}
\label{PD_U80}
\end{figure}
\par
Figures~\ref{chin_Um40_1} and \ref{chin_Um40_2} show the evolution of
the collective mode spectrum as imposed flow is increased.
As flow is increased, superfluidity is stable until one of the above instabilities occurs. Depending on filling and interaction strength, we can have different leading instabilities.
Taking into account all the instabilities and their critical momenta, we map
out a superflow `stability phase diagram' in the intermediate coupling regime
($U/t=4$) (Fig.~\ref{PD_U40}) and strong coupling regime ($U/t=8$)
(Fig.~\ref{PD_U80} left). 
We compare the strong coupling phase diagram obtained from GRPA with the one
obtained from spin-wave analysis in the strong coupling regime
(Fig.~\ref{PD_U80} right) - see Appendix C for details.
\par
In the intermediate and strong coupling regimes, the collective mode
spectrum always hits zero well before the two particle continuum does
(see Figs.~\ref{chin_Um40_1} and \ref{chin_Um40_2}). 
As a result, the critical flow
momentum is always set by the collective mode as shown in the phase diagram in Fig.~\ref{PD_U40}. 
The gapless superfluid phase and the depairing instability only exist at
mean field level, and will play no role in experiments. They may become
relevant to experiments at weak-coupling strengths. We discuss the
instabilities seen in the stability phase diagram below:
\par
\underline{Dynamical instability at $\Gamma$}:
In the vicinity of half-filling (see Fig.~\ref{chin_Um40_1}d), the Leggett mode has a small gap. This gap is a measure of the energy difference between the superfluid ground state and the low-lying sublattice-CDW phase. 
When a finite superflow is imposed, the kinetic energy of superflow overwhelms this excitation energy cost and the system switches to the CDW phase. This manifests itself as
the dynamical instability of the Leggett mode at the $\Gamma$ point. This is shown in Figs.~\ref{chin_Um40_1} (d)$\sim$(f). As flow is increased, the Leggett mode goes down in energy and hits
zero at the critical flow momentum. Subsequently, this mode energy
becomes complex and sublattice-CDW fluctuations grow exponentially.
We see this dynamical instability at the $\Gamma$ point in the vicinity
of and indeed precisely at half-filling (Fig.~\ref{chin_Um40_1} (a)$\sim$(c)).
\par
\underline{Landau instability}:
At low densities, since the system is similar to a continuum Fermi gas,
the leading superflow instability is naturally a Landau instability of
the AB mode.
With imposed flow, the AB mode undergoes a `Doppler shift' in the direction opposite
to flow and becomes negative at long wavelengths (see
panel (d)$\sim$(f) in Fig.~\ref{chin_Um40_2}). This type of instability
associated with negative energy collective modes has been observed in
superfluid fermions in shallow optical lattices leading to the onset of
dissipation \cite{Miller}.
\par
\underline{Dynamical instability at $M$}:
Surprisingly, at intermediate fillings ($n\simeq 0.4$), the leading
superflow instability is a dynamical instability of the Leggett mode at
the $M$ point (see panels (a)$\sim$(c) in Fig.~\ref{chin_Um40_2}). At low
densities, this feature persists as a subleading instability occurring
beyond the leading Landau instability (see Figs.~\ref{PD_U80}).
At low densities, the unstable mode is dominated by phase
fluctuations. At intermediate densities, it involves phase and density
fluctuations in roughly equal measure. With our choice of superflow
direction ($\bm Q$ along $\Gamma-K$ direction), this instability occurs
at two $M$ points which are symmetric with respect to the flow
direction. This instability occurs even when the flow is imposed along
the $\Gamma-M$ direction; in which case, only one $M$ point becomes
unstable. The nature of the system beyond this instability is an
exciting open question - we speculate that either local phase textures
will form with local superfluid currents, or that the system will become
chaotic.
\par
This $M$-point instability persists even in the strong coupling limit as
shown in Fig.~\ref{PD_U80}. This is in sharp contrast to the square
lattice case where only Landau and sublattice-CDW instabilities occur in
the strong coupling limit.  
In the strong coupling regime at low densities, we expect the physics to
be dominated by phase fluctuations, captured in a Quantum Rotor (QR)
model. In the square lattice case, the QR model is successful in
predicting the first \textit{dynamical} instability. This occurs at a
flow momentum of $\pi/2$ \cite{Polkovnikov} with the unstable wavevector
at the BZ edge center. On the honeycomb lattice, the QR
model predicts a dynamical instability at the flow momentum $\bm Q =
(\pi/\sqrt{3},0)$ with the unstable wavevector at the $M$ point. While it
gets the wavevector of the instability right, it overestimates the
critical flow momentum. Within GRPA and spin wave theory, we find that this
instability occurs at $\bm Q \sim (2\pi/3\sqrt{3},0)$.
Since density fluctuations are not included in the QR model, they may
suppress the critical flow momentum and give rise to the quantitative
discrepancy. To clarify this issue, it may be useful to experimentally 
study the critical flow momentum for non-interaction bosons on a honeycomb 
lattice. For the square lattice case, such an experiment has been carried 
out \cite{Mun} confirming the QR model result. 

\par
\underline{Incommensurate dynamical instability}:
In Fig.~\ref{PD_U40}, the leading instability for
intermediate fillings ($0.5\lesssim n\lesssim 0.7$) at $U/t=4$ is an
incommensurate dynamical instability. This instability can be seen in
Fig.~\ref{chin_Um40_1} (g)$\sim$(i) with the unstable wavevector between
the $\Gamma$ and $M$ points. Upon decreasing filling, the unstable
wavevector moves towards the $M$ point.
Such an instability does not occur in the strong-coupling regime as shown in Fig.~\ref{PD_U80}.
This intermediate-coupling incommensurate instability has also been seen
in square/cubic lattices \cite{Ganesh,Yunomae}. 
\par
Figure~\ref{PD_U80}(right) shows the phase diagram in the strong-coupling limit from
the spin wave theory. It exhibits qualitatively the same features as 
the strong-coupling phase diagram obtained by GRPA in
Fig.~\ref{PD_U80}(left): dynamical instability of the Leggett mode near
the $\Gamma$ point around half-filling, Landau instability 
of the AB mode at low density, and the dynamical instability of the
Leggett mode at the $M$ points for intermediate fillings. This demonstrates the consistency of the results obtained 
from GRPA and from spin wave analysis, serving as a check for the validity of GRPA.
\par
In addition to the superflow instability discussed above, we point out 
some characteristic features of the spectra in
Figs.~\ref{chin_Um40_1} and \ref{chin_Um40_2}. Close to half-filling,
the Leggett mode is dominated by density fluctuations, while the
AB mode is predominantly composed of superfluid phase fluctuations. This
is clearly seen in Fig.~\ref{chin_Um40_1} (e) and (h) from the small level
repulsion between the two collective modes in the $\Gamma-M$ region.
At half-filling, as seen in Fig.~\ref{chin_Um40_1} (e) and (h), two branches of the collective mode overlap, indicating that their fluctuation components are orthogonal.
On the other hand, at low fillings, since both collective
modes are composed of superfluid phase fluctuations, they exhibit large level repulsion
in Fig.~\ref{chin_Um40_2} (b) and (h) in the $\Gamma-M$ region.
We also note that, at low densities, the Leggett mode enters the two
particle continuum when a small flow is imposed as shown in
Fig.~\ref{chin_Um40_2}. Consequently, the peaks
become broad as they are damped by decay into single-particle excitations. However, the
collective mode peaks still possess high intensity and appear as
bright regions inside the continuum.

\section{Summary and Discussion}
\label{conclusions}

We have studied the $s$-wave superfluid state on the honeycomb lattice inspired by recent cold atom experiments. 
The optical lattice setup of Ref.~\onlinecite{Tarruell} realizes a honeycomb geometry with tunable anisotropy. Starting from this configuration, we deduce that the symmetric lattice limit is optimal for the study of superfluidity. 
In addition, superfluidity is most robust at fillings close to $n=3/4,5/4$ fermions per site, due to a vH singularity in the non-interacting problem - this is the optimal density for observing superfluidity. 
Indeed, we find that the highest mean field $T_c^0$ occurs at the
density corresponding to the vH singularity and in the symmetric lattice
case. 
\par
The critical velocity of superflow is a powerful tool to understand the
low-lying excitations of a superfluid. To study this quantity, we first
perform a mean field analysis of the superfluid state while allowing for imposed superflow. Experimentally, superflow can be imposed by using a `running' optical lattice potential. 
On the honeycomb lattice, we find that superflow generically drives the
system into a gapless superfluid state which is widest in the vicinity
of the vH singularity at $n=3/4,5/4$. Subsequently, the mean field order
parameter vanishes in a first order depairing transition. 
\par
We go beyond mean field theory by including density and pairing
fluctuations using the GRPA formalism. At strong coupling, we compare
our GRPA results with a spin-wave analysis of the appropriate pseudospin
Hamiltonian. Although strictly justified only in the weak-coupling
limit, GRPA works well even at strong coupling.
We obtain sharp collective mode excitations with two branches - an
Anderson-Bogoliubov mode and a Leggett mode. The former incorporates the
Nambu-Goldstone mode from U(1) symmetry breaking. The latter is composed of
out-of-phase fluctuations between the two sites of the honeycomb unit
cell. At half-filling in the stationary superfluid, this Leggett mode
becomes gapless at the $\Gamma$ point, due to an extra pseudospin
symmetry of the Hubbard model. This arises due to the sublattice-CDW
state becoming degenerate with the superfluid. Away from half-filling,
the Leggett mode acquires a gap reflecting the energy difference between
the superfluid and the CDW states.
\par
With imposed superflow, the collective modes come down in energy due to
a Doppler shift. An instability occurs when the collective mode softens
- subsequently, the mode either becomes negative (Landau instability) or
complex-valued (dynamical instability). Close to half-filling, the
leading instability is a dynamical instability of the Leggett mode at
the $\Gamma$ point. This instability arises due to competition between
the superfluid and sublattice-CDW phases. The presence of the low-lying
CDW phase is a special feature of the one-band Hubbard model, arising
from a delicate pseudospin rotational symmetry at half-filling. For
future experiments, the presence of this instability close to
half-filling can serve as a non-trivial test to verify if the one-band
Hubbard model has indeed been realized. Na\"ively, we may expect this
instability to give rise to a stable `supersolid' phase with coexisting
superfluid and CDW orders. However, as in the square lattice case \cite{Burkov, Ganesh}, the honeycomb system is likely to go into a time-dependent non-equilibrium state. To probe this instability in experiments, it is best to focus on the short time window right after the critical momentum is reached. The exponentially growing CDW correlations can then be probed by rapidly ramping up the optical lattice to freeze the fermion density at each site. Subsequent imaging should give a snapshot clearly showing frozen CDW correlations. Even as we approach the critical flow momentum from below, we can use the same procedure to obtain snapshots which will show CDW fluctuations over some characteristic length scale. This length scale should diverge as we approach the critical flow momentum.
\par
At low densities, we find a Landau instability of the AB mode. We also find an unexpected dynamical instability at the M point for intermediate densities. Surprisingly, this instability persists even at strong coupling. The nature of the system beyond this instability is an interesting open question. We present stability phase diagrams which give the critical flow momentum and the associated instability mechanism as a function of density and interaction strength. 
\par
Our calculations are also relevant to recent studies of the
\textit{repulsive} Hubbard model on the honeycomb lattice. At
half-filling, quantum Monte Carlo simulations \cite{Meng} have suggested a
gapped spin liquid state at intermediate $U/t$  between semi-metal and
the N\'eel phases. There have been many attempts to understand this
intermediate phase and the nature of phase transitions into and out of
it. 
Precisely at half-filling, the repulsive and attractive Hubbard
models are related by a sublattice-dependent particle hole
transformation. In fact, the attractive model at any filling is
equivalent to the repulsive problem at half-filling, but with a magnetic
field \cite{Giamarchi}. Our GRPA approach does not capture a
strongly correlated spin-liquid phase as suggested by Quantum Monte Carlo simulations. 
However, approaching the spin
liquid from the N\'eel side, the N\'eel-spin liquid quantum phase
transition may arise from melting of N\'eel order due to collective mode
excitations. Such a mechanism has been discussed in
Ref.\onlinecite{Sheng} in the context of the Bi$_3$Mn$_4$O$_{12}$NO(3),
which is a spin-3/2 honeycomb lattice antiferromagnet. The question can be
phrased within the attractive Hubbard model - can collective modes
obtained from GRPA melt superfluid order? If so, what is the nature of
the resulting state? 
An exciting possibility is to obtain a `pairing liquid' state which is
the analog of the spin liquid proposed for the repulsive
model \cite{YingRan}. Such a pairing liquid can be thought of as a \textit{quantum}
pseudogap phase with preformed Cooper pairs. Recently, a
finite-temperature pseudogap phase has been observed in a trapped Fermi
gas near the unitary limit \cite{Gaebler,Tsuchiya}. However, a zero-temperature analog has not
been explored. This is an interesting direction for future research.

\acknowledgements
S. T. thanks T. Esslinger, M. Sigrist, T. Nikuni, Y. Ohashi,
M. Marmorini, D. Yamamoto, S. Konabe, and I. Danshita for
fruitful discussions, and acknowledges L. Tarruell, D. Greif, and
T. Uehlinger for valuable information about the experiments.
S. T. was supported by Grant-in-Aid for
Scientific Research, No. 24740276. A. P. was supported by NSERC of Canada.

\appendix
\section{Mean field theory equations}
\label{app.MFTeqs}
The Hamiltonian of Eq.~(\ref{Hflow}) can be solved by a Bogoliubov transformation as given in Eq.~(\ref{BogTrans}).
Denoting the elements of the transformation matrix $\hat{T}$ by $(\hat T)_{ij}=T_{ij}$,
the gap equation (\ref{gap}) gives
\begin{eqnarray}
\Delta_{\bm Q,a}&=&\frac{U}{M}\sum_{\bm k}\left[T_{21}^\ast  T_{11}f(E^+_{\bm Q}(\bm k))+T_{22}^\ast
					   T_{12}\left(1-f(E^+_{\bm Q}(-\bm
					   k))\right)\right.\nonumber\\
&&\left.+T_{23}^\ast T_{13}f(E^-_{\bm Q}(\bm k))+T_{24}^\ast
					   T_{14}\left(1-f(E^-_{\bm Q}(-\bm
					   k))\right)\right],\label{gapA}\\
\Delta_{\bm Q,b}&=&\frac{U}{M}\sum_{\bm k}\left[T_{41}^\ast  T_{31}f(E^+_{\bm Q}(\bm k))+T_{42}^\ast
					   T_{32}\left(1-f(E^+_{\bm Q}(-\bm
					   k))\right)\right.\nonumber\\
&&\left.+T_{43}^\ast T_{33}f(E^-_{\bm Q}(\bm k))+T_{44}^\ast
					   T_{34}\left(1-f(E^-_{\bm Q}(-\bm
					   k))\right)\right],\label{gapB}
\end{eqnarray}
where $\langle \alpha_{\bm k,\tau,\sigma}^\dagger\alpha_{\bm
k,\tau,\sigma}\rangle=f(E^\tau_{\bm Q}(\bm k))$ ($f(E)\equiv 1/\{e^{\beta E}+1\}$ is the Fermi distribution function).

The average filling per site $n$ is 
\begin{eqnarray}
n&=&\frac{1}{N}\sum_{\bm k}\sum_{\nu,\sigma}\langle n_{\bm k,\nu,\sigma}\rangle\nonumber\\
&=&\frac{1}{N}\sum_{\bm k}\left[ \left(|T_{11}|^2-|T_{21}|^2+|T_{31}|^2-|T_{41}|^2\right)f(E^+_{\bm Q}(\bm k))\right.\nonumber\\
&&+\left(-|T_{12}|^2+|T_{22}|^2-|T_{32}|^2+|T_{42}|^2\right)f(E^+_{\bm Q}(-\bm k))\nonumber\\
&&+\left(|T_{13}|^2-|T_{23}|^2+|T_{33}|^2-|T_{43}|^2\right)f(E^-_{\bm Q}(\bm k))\nonumber\\
&&+\left(-|T_{14}|^2+|T_{24}|^2-|T_{34}|^2+|T_{44}|^2\right)f(E^-_{\bm Q}(-\bm k))\nonumber\\
&&\left.+|T_{12}|^2+|T_{14}|^2+|T_{21}|^2+|T_{23}|^2+|T_{32}|^2+|T_{34}|^2+|T_{41}|^2+|T_{43}|^2\right].
\label{numbereq}
\end{eqnarray}
With imposed superflow ($\bm Q \neq 0$), these equations have to be
solved numerically. For the stationary superfluid, however, setting
$\Delta_{\bm Q,a}=\Delta_{\bm Q,b}=\Delta_0$, the $\hat{T}$ matrix can be obtained analytically.
\begin{equation}
\hat{T}_{\bm Q = 0}=\frac{1}{\sqrt{2}}
\left(
\begin{array}{cccc}
e^{i\phi_{\bm k}}u_{\bm k}^+ & e^{i\phi_{\bm k}}v_{\bm k}^+ & u_{\bm k}^- & v_{\bm k}^- \\
-e^{i\phi_{\bm k}}v_{\bm k}^+ & e^{i\phi_{\bm k}}u_{\bm k}^+ & -v_{\bm k}^- & u_{\bm k}^- \\
u_{\bm k}^+ & v_{\bm k}^+ & -e^{-i\phi_{\bm k}}u_{\bm k}^- & -e^{-i\phi_{\bm k}}v_{\bm k}^- \\
-v_{\bm k}^+ & u_{\bm k}^+ & e^{-i\phi_{\bm k}}v_{\bm k}^- & -e^{-i\phi_{\bm k}}u_{\bm k}^-
\end{array}
\right),
\label{T}
\end{equation}
where $e^{i\phi_{\bm k}}=\gamma_{\bm k}/|\gamma_{\bm k}|$. 
The matrix elements $u_{\bm k}^\pm$ and $v_{\bm k}^\pm$ in Eq.~(\ref{T}) have the
standard form as
\begin{eqnarray}
u_{\bm k}^\pm=\sqrt{\frac{1}{2}\left(1+\frac{\xi_{\bm k}^\pm}{E^\pm(\bm
			k)}\right)}~,\ \ 
v_{\bm k}^\pm=\sqrt{\frac{1}{2}\left(1-\frac{\xi_{\bm k}^\pm}{E^\pm(\bm k)}\right)}~.
\end{eqnarray}
\par
The gap equation (\ref{gapA}) and (\ref{gapB}) reduce to 
\begin{eqnarray}
\Delta_0
&=&\frac{U}{N}\sum_{\bm k}\sum_{\tau=\pm}u_{\bm k}^\tau v_{\bm
					     k}^\tau\left[1-2f(E^\tau(\bm k))\right]\nonumber\\
&=&\frac{U\Delta_0}{N}\sum_{\bm k}\sum_{\tau=\pm}\frac{\tanh\frac{\beta
 E^\tau(\bm k)}{2}}{2E^\tau(\bm k)}.
\end{eqnarray}
Thus, we obtain
\begin{equation}
\frac{1}{U}=\frac{1}{N}\sum_{\bm k}\sum_{\tau=\pm}\frac{\tanh\frac{\beta
 E^\tau(\bm k)}{2}}{2E^\tau(\bm k)}.
\label{gapeq}
\end{equation}
The number equation (\ref{numbereq}) becomes
\begin{eqnarray}
n&=&\frac{1}{M}\sum_{\bm k}\sum_{\tau=\pm}\left[\{(u_{\bm k}^\tau)^2-(v_{\bm k}^\tau)^2\}f(E^\tau(\bm
		 k))+(v_{\bm k}^\tau)^2\right]\nonumber\\
&=&1-\frac{1}{N}\sum_{\bm k}\sum_{\tau=\pm}\frac{\xi_{\bm k}^\tau}{E^\tau(\bm
		   k)}\tanh\frac{\beta E^\tau(\bm k)}{2}.
\label{number}
\end{eqnarray}
The above equations (\ref{gapeq}) and (\ref{number}) have been derived
in Ref.~\cite{Zhao}.

\section{GRPA formalism of C\^ot\'e and Griffin}
\label{app.GRPA}
In this Appendix, we give a brief summary of the GRPA formalism
developed by C\^ot\'e and Griffin \cite{Cote}.
In this formalism, we introduce fictitious external fields to calculate various response functions by taking
functional derivatives of the single-particle Green's function with respect
to the external fields. The collective mode spectrum of the
system is given by the poles of the response functions.
As emphasized in Ref.~\cite{Cote}, this technique has the advantage that
the response functions are uniquely determined by the functional
differentiation technique to be consistent with the
self-energy approximation, and therefore gives a {\it gapless} Nambu-Goldstone mode in the
superfluid phase.
\par
The Hamiltonian with time-dependent external fields coupled to density and
pair operators is
\begin{equation}
K(\tau)=H+\sum_i P_i(\tau)n_i+\sum_i\left(Q^\ast_i(\tau)c_{i\uparrow}c_{i\downarrow}+Q_i(\tau)c^\dagger_{i\downarrow}c^\dagger_{i\uparrow}\right),
\end{equation}
where $\tau$ is imaginary time. Hereafter, we adopt the notation $1\equiv(j_1,\tau_1)=(l_1,\nu_1,\tau_1)$, where $j_1$
denotes the lattice site, while $l_1$ and $\nu_1$ denote the unit cell and
sublattice, respectively.
It is convenient to introduce the Nambu representation \cite{Nambu}, in
which a single-particle Green's function is given in a $2\times 2$ matrix form,
\begin{eqnarray}
\hat G(1,2)&=&-\langle T \Psi(1)\Psi^\dagger(2)\rangle\nonumber\\
&=&\left(
\begin{array}{cc}
-e^{-i\frac{\bm Q}{2}\cdot\bm(\bm r_{l_1}-\bm r_{l_2})}\langle Tc_\uparrow(1)c^\dagger_\uparrow(2)\rangle & -e^{-i\frac{\bm
 Q}{2}\cdot(\bm r_{l_1}+\bm r_{l_2})}\langle T
 c_\uparrow(1)c_\downarrow(2)\rangle \\
-e^{i\frac{\bm Q}{2}\cdot(\bm r_{l_1}+\bm r_{l_2})}\langle T c_\downarrow^\dagger(1)c^\dagger_\uparrow(2)\rangle & e^{i\frac{\bm
 Q}{2}\cdot(\bm r_{l_1}-\bm r_{l_2})}\langle Tc_\downarrow(2)c^\dagger_\downarrow(1)\rangle \\
\end{array}
\right).
\label{Greensfunc}
\end{eqnarray}
The field operator $\Psi(1)$ is a spinor consisting of the creation and annihilation operators
\begin{equation}
\Psi(1)=\hat \gamma_{l_1}
\left(
\begin{array}{c}
c_{\uparrow}(1)\\
c^\dagger_{\downarrow}(1)
\end{array}
\right),
\Psi^\dagger(1)=\left(c_{\uparrow}^\dagger(1),c_{\downarrow}(1)\right)\hat\gamma_{l_1}^*.
\end{equation}
The unitary matrix $\hat \gamma_{l_1}$ is given by
\begin{equation}
\hat\gamma_{l_1}=
\left(
\begin{array}{cc}
e^{-i\frac{\bm Q}{2}\cdot\bm r_{l_1}} & 0 \\
0 & e^{i\frac{\bm Q}{2}\cdot\bm r_{l_1}}
\end{array}
\right).
\end{equation}
We note that the Green's function in Eq.~(\ref{Greensfunc}) is defined in the
co-moving frame of the superfluid. The Green's function in the laboratory frame is
obtained by the unitary transformation $\hat\gamma_{l_1}^*\hat G(1,2)\hat\gamma_{l_2}$.
\par
The Green's function $\hat G(1,2)$ satisfies the Dyson equation 
\begin{eqnarray}
[\hat G(1,2)]^{-1}=[\hat G^0(1,2)]^{-1}-\hat\Sigma(1,2)-\hat W(1,2),
\end{eqnarray}
where $\hat G^0(1,2)$ is the non-interacting Green's function.
The external field matrix $\hat W(1,2)$ is defined as
\begin{equation}
\hat W(1,2)\equiv
\delta(1-2)\left(
\begin{array}{cc}
P(1) & -e^{-i\bm Q\cdot\bm r_{l_1}}Q(1) \\
-e^{i\bm Q\cdot\bm r_{l_1}}Q^*(1) & -P(1)
\end{array}
\right).
\end{equation}
The self-energy within the Hartree-Fock-Gor'kov (HFG) approximation is
given by
\begin{equation}
\hat\Sigma(1,2)=
\delta(1-2)\left(
\begin{array}{cc}
-Un/2 & -\Delta_{\bm Q,\nu_1} \\
-\Delta_{\bm Q,\nu_1}^* & Un/2
\end{array}
\right).
\end{equation}
\par
We introduce the three-point {\it density} correlation function as
\begin{equation}
\hat L(1,2,3)\equiv\frac{\delta\hat{\bar G}(1,2) }{\delta
 P(3)},
\label{defL}
\end{equation}
where $\hat{\bar G}(1,2)\equiv\hat\tau_3\hat G(1,2)$ with $\tau_3$ being the usual Pauli matrix. In the
equal space-time limit, it reduces to
\begin{equation}
\hat L(1,3)\equiv\hat L(1,1^+,3)=
\left(
\begin{array}{cc}
-\langle T\delta n_\uparrow(1)\delta n(3)\rangle & -e^{-i\bm Q\cdot\bm
 r_{l_1}}\langle T\delta m(1)\delta n(3)\rangle \\
e^{i\bm Q\cdot\bm r_{l_1}}\langle T\delta m^\dagger(1)\delta n(3)\rangle
 & -\langle T\delta n_\downarrow(1)\delta n(3)\rangle
\end{array}
\right).
\label{equalL}
\end{equation}
The density response function is thus obtained from the diagonal
elements of Eq.~(\ref{equalL}) as
\begin{eqnarray}
\chi_d(1,3)&=&\frac{\delta\langle n(1)\rangle}{\delta
 P(3)}=-\langle T\delta n(1)\delta n(3)\rangle\nonumber\\
&=&L_{11}(1,3)+L_{22}(1,3).
\label{chid}
\end{eqnarray}
We also introduce the three-point {\it pair} correlation function as
\begin{equation}
\hat M(1,2,3)\equiv e^{i\bm Q\cdot\bm r_{l_3}}\frac{\delta \hat{\bar G}(1,2)}{\delta
 Q(3)}.
\end{equation}
In the equal space-time limit, it reduces to
\begin{equation}
\hat M(1,3)\equiv\hat M(1,1^+,3)=
\left(
\begin{array}{cc}
-e^{i\bm Q\cdot\bm r_{l_3}}\langle T\delta n_\uparrow(1)\delta m^\dagger(3)\rangle & -e^{-i\bm Q\cdot(\bm
 r_{l_1}-\bm r_{l_3})}\langle T\delta m(1)\delta m^\dagger(3)\rangle \\
e^{i\bm Q\cdot(\bm r_{l_1}-\bm r_{l_3})}\langle T\delta m^\dagger(1)\delta
 m^\dagger(3)\rangle & -e^{i\bm Q\cdot\bm r_{l_3}}\langle T\delta n_\downarrow(1)\delta m^\dagger(3)\rangle
\end{array}
\right).\label{pairM}
\end{equation}
The pair response function is obtained from the off-diagonal element
of Eq.~(\ref{pairM}) as
\begin{eqnarray}
\chi_\Delta(1,3)&=&e^{-i\bm Q\cdot(\bm r_{l_1}-\bm r_{l_3})}\frac{\delta\langle
 m(1)\rangle}{\delta Q(3)}\nonumber\\
&=&-e^{-i\bm Q\cdot(\bm r_{l_1}-\bm
 r_{l_3})}\langle T\delta m(1)\delta m^\dagger(3)\rangle
= M_{12}(1,3).
\label{chidelta}
\end{eqnarray}
\par
\begin{figure}
\centerline{\includegraphics[width=10cm]{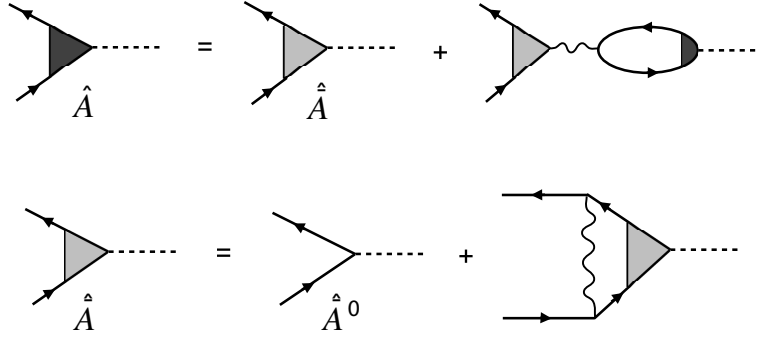}}
\caption{The diagrammatic representation of the GRPA equations
 (\ref{GRPAladder}) and (\ref{GRPAbubble}). The three
 point correlation function $\hat A$ is either the
 density ($\hat L$) or pair ($M$) correlation function. 
The wavy line represents the attractive interaction $-U$.
 $\hat A$ involves the bubble diagrams, while the irreducible
 correlation function $\hat{\bar A}$ involves the ladder diagrams.
}
\label{GRPA_diagram}
\end{figure}

Differentiating $\hat G(1,2)$ with respect to the external
fields and using the HFG approximation, one obtains the GRPA equation of motion for the
three-point response functions as
\begin{eqnarray}
\hat A(1,2,5)&=&\hat{\bar A}(1,2,5)-U\int d3\ \hat{\bar
 L}(1,2,3)[A_{11}(3,5)+A_{22}(3,5)],\label{GRPAladder}\\
\hat{\bar A}(1,2,5)&=&\hat A^0(1,2,5)+U\int d3\ \hat{\bar G}(1,3)\hat{\bar
 A}(3,5)\hat{\bar G}(3,2),
\label{GRPAbubble}
\end{eqnarray}
where $\hat A$ is either $\hat L$ or $\hat M$. 
The lowest-order correlation
functions are given by
\begin{eqnarray}
\hat L^0(1,2,5)&\equiv&\hat{\bar G}(1,5)\hat{\bar G}(5,2),\\
\hat M^0(1,2,5)&\equiv& \hat{\bar G}(1,5)
\left(
\begin{array}{cc}
0 & 1 \\
0 & 0
\end{array}
\right)\hat{\bar G}(5,2).
\label{M0}
\end{eqnarray}
$\hat{\bar A}(1,2,3)$ is called the irreducible correlation function. 
The form of Eqs.~(\ref{GRPAladder}) and (\ref{GRPAbubble}) exposes the
diagrammatic structure of the GRPA, as shown in Fig.~\ref{GRPA_diagram},
where Eqs.~(\ref{GRPAladder}) and (\ref{GRPAbubble}) represent the
summation of the ladder diagrams and bubble diagrams, respectively.

\par
The GRPA equations in momentum space are
\begin{eqnarray}
\hat{A}^{\nu_1\nu_2}(q)&=&\hat{\bar A}^{\nu_1\nu_2}(q)-U\sum_{\nu_3}\hat{\bar L}^{\nu_1\nu_3}(q)A^{\nu_3\nu_2}(q),\label{GRPA1}\\
\hat{\bar A}^{\nu_1\nu_2}(q)&=&\hat A^{0\nu_1\nu_2}(q)+\frac{U}{\beta
 M}\sum_{\nu_3}\sum_{\bm p,\omega_n}\hat{\bar G}^{\nu_1\nu_3}(p)
\hat{\bar A}^{\nu_3\nu_2}(q)\hat{\bar
 G}^{\nu_3\nu_1}(p-q),\label{GRPA2}
\end{eqnarray}
where $q\equiv(\bm q,i\Omega_n)$ and $p\equiv(\bm p,i\omega_n)$
($\omega_n$ and $\Omega_n$ are Fermi and Bose Matsubara frequencies, respectively). Here, we use the Fourier transform
\begin{eqnarray}
\hat{\bar G}^{\nu_1\nu_2}(1,2)&=&\frac{1}{\beta M}\sum_{\bm
 p,\omega_n}\hat{\bar G}^{\nu_1\nu_2}(p)e^{i(\bm p\cdot(\bm
 r_{l_1}-\bm r_{l_2})-\omega_n(\tau_1-\tau_2))},\\
\hat A^{\nu_1\nu_2}(1,2)&=&\frac{1}{\beta M}\sum_{\bm q,\Omega_n}\hat
 A^{\nu_1\nu_2}(q)e^{i(\bm q\cdot(\bm
 r_{l_1}-\bm r_{l_2})-\Omega_n(\tau_1-\tau_2))}.
\label{fourier}
\end{eqnarray}
Using Eqs.~(\ref{chid}) and (\ref{chidelta}), the density and pair response functions can be evaluated from the
correlation functions $\hat L$ and $\hat M$ that are obtained by solving
Eqs.~(\ref{GRPA1}) and (\ref{GRPA2}). 

\section{Spin-wave analysis in the strong coupling limit}
\label{app.spinwave}
In the limit of large U/t, states with singly occupied sites can be
ignored leading to an effective spin-1/2 Heisenberg model
\cite{Ganesh}. The effective Hamiltonian can be written as 
\begin{eqnarray}
H = J \sum_{i,\bdelta=0,a_2,a_1+a_2} \bS_{i,a} \cdot \bS_{i+\bdelta,b} - (BS)\sum_{i,\nu=a,b} S_{i,\nu}^{z}
\end{eqnarray}
The exchange interaction $J=4t^2/U$ arises from superexchange. In this spin mapping, the X and Y directions of spin correspond to the local superfluid order parameter. The density of fermions maps onto the z-polarization, with half-filling mapping to zero z-polarization. The chemical potential in the fermion problem gives rise to a magnetic field tuning polarization. We take the magnetic field to be proportional to S, so that all terms in the Hamiltonian scale as $S^2$. 

The uniform superfluid state at half-filling maps onto a N\'eel antiferromagnetic order with spins lying in the XY plane. This state is symmetric under rotation about the Z axis, which corresponds to the usual U(1) gauge symmetry. Superflow is imposed by a phase gradient, which corresponds to twisting the local N\'eel vector at by an angle $\bQ\cdot \br_i$ leading to a spiral state with spins lying in the XY plane. Away from half-filling, a non-zero magnetic field is introduced. This forces the spins to cant uniformly towards the Z axis to gain Zeeman energy. We characterize the flowing superfluid as
\begin{eqnarray}
\nn \bS_{i,a} = S(\cos\chi\cos(\bQ.\br_i),\cos\chi\sin(\bQ.\br_i),\sin\chi)\\
\bS_{i,b} = S(-\cos\chi\cos(\bQ.\br_i+\phi),-\cos\chi\sin(\bQ.\br_i+\phi),\sin\chi)
\end{eqnarray}
We allow the A and B sublattices to have a relative phase difference $\phi$. The N\'eel state (stationary superfluid) is recovered by setting $\bQ=\phi=0$. For simplicity, we denote the phase angles by $\alpha_i=\bQ\cdot\br_i$ and $\beta_i=\bQ\cdot\br_i+\phi$. 
We now perform a local spin rotation and define new spin operators which we denote as ${\mathbf T}_{i,\nu}=W_{i,\nu} \bS_{i,\nu}$, by 
\begin{eqnarray}
\left(\begin{array}{c}
       T_{i,a}^x \\ T_{i,a}^y \\ T_{i,a}^z
      \end{array} \right) =
    \left(\begin{array}{ccc}
        \sin\chi & 0 & -\cos\chi \\
        	0 & 1 & 0 \\
       \cos\chi & 0 & \sin\chi 
 \end{array}\right)
    \nn  \left(\begin{array}{ccc}
    \cos(\alpha_i) & \sin(\alpha_i) & 0 \\
    -\sin(\alpha_i) & \cos(\alpha_i) & 0 \\
    	0 & 0 & 1 \end{array}\right)
\left(\begin{array}{c}
       S_{i,a}^x \\ S_{i,a}^y \\ S_{i,a}^z
      \end{array} \right),
\end{eqnarray}
\begin{eqnarray}
\left(\begin{array}{c}
       T_{i,b}^x \\ T_{i,b}^y \\ T_{i,b}^z
      \end{array} \right) =
\left(\begin{array}{ccc}
	-\sin\chi & 0 & -\cos\chi\\
	0 & 1 & 0 \\
	\cos\chi & 0 & -\sin\chi    \end{array}\right) 
    \left( \begin{array}{ccc}    
    \cos(\beta_i) & \sin(\beta_i) & 0 \\
    -\sin(\beta_i) & \cos(\beta_i) & 0 \\
    	0 & 0 & 1 \end{array}\right)
\left(\begin{array}{c}
       S_{i,b}^x \\ S_{i,b}^y \\ S_{i,b}^z
      \end{array} \right).
\end{eqnarray}
With this local spin rotation, the flowing superfluid state transforms into a uniform antiferromagnet with all spins pointing towards $\pm$Z. We absorb the spin rotation operators into the Hamiltonian. The exchange coupling then takes the form 
\begin{eqnarray}
\nn \bS_{i,a}\cdot\bS_{j,b}
=\left( \begin{array}{c}
T_{i,a}^{x} \\ T_{i,a}^{y} \\ T_{i,a}^z \end{array}\right)^{T}
\left(\begin{array}{ccc}
C_\chi^2 - S_\chi^2 C_{\alpha_i-\beta_j} & S_\chi 
S_{\alpha_i-\beta_j} & S_{2\chi}C_{(\alpha_i-\beta_j)/2}^2 \\
S_\chi S_{\alpha_i-\beta_j} & C_{\alpha_i-\beta_j} & -C_\chi S_{\alpha_i-\beta_j} \\
-S_{2\chi} C_{(\alpha_i-\beta_j)/2}^2 & C_\chi S_{\alpha_i-\beta_j} & 
C_\chi^2 C_{\alpha_i-\beta_j} - S_\chi^2 \end{array}\right)
\left( \begin{array}{c}
T_{j,b}^{x} \\ T_{j,b}^{y} \\ T_{j,b}^z \end{array}\right),
\end{eqnarray}
where $j=i+\bdelta$. We denote $\sin\theta \equiv S_\theta$ and $\cos\theta \equiv C_\theta$. The Zeeman term in the Hamiltonian becomes
\begin{eqnarray}
-BS \left( S_{i,a}^{z} + S_{i,b}^{z} \right) = -BS \left( -C_\chi \{ T_{i,a}^x + T_{i,b}^x \} + S_\chi \{ T_{i,a}^z - T_{i,b}^z \} \right).
\end{eqnarray}

We now introduce Holstein Primakov bosons as follows: \\
\begin{center}
\begin{tabular}{ll}
$T_{i,a}^{x} = \sqrt{\frac{S}{2}}(a_i+a_i^\dg) $, & 
$\phantom{abc}T_{i,b}^{x} = \sqrt{\frac{S}{2}}(b_i+b_i^\dg)$,  \\
$T_{i,a}^{y} = \sqrt{\frac{S}{2}}\left(\frac{1}{i}\right)(a_i-a_i^\dg)$, & 
$\phantom{abc}T_{i,b}^{y} = -\sqrt{\frac{S}{2}}\left(\frac{1}{i}\right)(b_i-b_i^\dg)$, \\
$T_{i,a}^{z} = S-a_{i}^\dg a_i$, & 
$\phantom{abc}T_{i,b}^{z} = b_i^\dg b_i - S$.\\
\end{tabular}\end{center}
The bosonic operators $a_i$ and $b_i$ represent fluctuations about the spiral state. S is the spin length, which we will ultimately set to be $S=1/2$. The Hamiltonian can now be expanded in powers of S. The leading terms, of order $S^2$, give the energy in the classical limit. 
\begin{eqnarray}
E_{Cl} =2NS^2 \left[ -J C_\chi^2 \frac{\gamma_{\bQ}e^{i\phi} + \gamma_{-\bQ}e^{-i\phi}}{2} + J S_\chi^2 \epsilon_0 
- 2BS_\chi \right],
\end{eqnarray}
where $N$ is the total  number of sites. We have denoted $\gamma_\bk = 1 + e^{i\bk \cdot a_2} + e^{i\bk \cdot (a_1+a_2)} $.
Terms proportional to $S^{3/2}$ are linear in boson operators.
The coefficient of the single boson operators is complex-valued - we set this quantity to be zero. 
Setting the real part to be zero, we get
\begin{eqnarray}
\sin\chi =  \frac{B}{J (3+ 
\frac{\gamma_\bQ e^{i\phi} + \gamma_{-\bQ}e^{-i\phi}}{2} )}.
\end{eqnarray}
Alternatively, this equation be obtained by treating $\chi$ as a variational parameter and demanding $\partial_{\chi}{E_{Cl}}=0$.
Setting the imaginary part of the coefficient to zero gives
\begin{eqnarray}
\tan \phi =  \frac{\sin(Q_2)+\sin(Q_1+Q_2)}{1+\cos(Q_2)+\cos(Q_1+Q_2)},
\end{eqnarray}
where $Q_1$ and $Q_2$ are the components of $\bm Q$ along the primitive vectors $\bm a_1$ and $\bm a_2$ shown in Fig.~\ref{fig.asymcouplings}. This fixes the value of $\phi$. Again, this equation can be also obtained variationally by demanding $\partial_{\phi}{E_{Cl}}=0$. 

The next terms in the Holstein-Primakov expansion are of order S. These capture the quantum spin wave fluctuations.
These terms are given by
\begin{eqnarray}
\nn H_{qu} = 2NS \left[3J S_\chi^2  
- J C_\chi^2 \frac{\gamma_\bQ e^{i\phi} + \gamma_{-\bQ}e^{-i\phi}}{2}
- B S_\chi \right]
+S \sum_{k} {}^{'} \psi_{\bk}^\dg H_{\bk} \psi_{\bk},
\end{eqnarray}
where
\begin{eqnarray}
\psi_{\bk} = \left(\begin{array}{c}
       a_{\bk} \\ a_{-\bk}^\dg \\ b_{\bk} \\ b_{-\bk}^\dg
      \end{array}\right);
H_{\bk}=\left(\begin{array}{cccc}
U_{\bk} & 0 & X_{\bk}+Y_{\bk} & Z_{\bk}\\
  & U_{\bk} & Z_{\bk} & X_{\bk}-Y_{\bk} \\
& & U_{\bk} & 0 \\
 &  & & U_{\bk}
      \end{array}\right).
\end{eqnarray}
As the Hamiltonian matrix is Hermitian, we have only filled in the upper triangle. The entries are given by 
\begin{eqnarray}
\nn U_{\bk} &=& J ( -3S_\chi^2  + C_\chi^2 
\frac{\gamma_\bQ e^{i\phi}+\gamma_{-\bQ}e^{-i\phi}}{2} ) + B S_\chi, \\
X_{\bk} &=& \frac{J}{2}\left[  C_\chi^2 \gamma_{\bk} - (1+S_\chi^2) \frac{\gamma_{\bk+\bQ}e^{i\phi}+\gamma_{\bk-\bQ}e^{-i\phi}  }{2}
\right], \\
Y_{\bk} &=& -J S_\chi \frac{\gamma_{\bk+\bQ}e^{i\phi}-\gamma_{\bk-\bQ}e^{-i\phi}}{2}, \\
Z_{\bk} &=& \frac{J}{2} C_\chi^2 \left[ \gamma_\bk +
\frac{\gamma_{\bk+\bQ}e^{i\phi}+\gamma_{\bk-\bQ}e^{-i\phi} }{2} \right].
\end{eqnarray}
We obtain the spin wave spectrum by performing a bosonic Bogoliubov transformation on this Hamiltonian matrix. For a given filling and flow momentum, we can thus obtain the collective mode excitations in the strong coupling limit.

\end{document}